\documentclass{mn2e}
\usepackage{epsfig,color}
\title{Insight into the OH polarimetric structure of OH~26.5+0.6}

\author[Etoka \& Diamond]
{S. Etoka$^1$\thanks{E-mail: Sandra.Etoka@googlemail.com},
P. Diamond$^1$\\
$^1$Jodrell Bank Centre for Astrophysics,
School of Physics and Astronomy, The University of Manchester, 
    Manchester M13 9PL, UK \\
}
\begin{document}
\maketitle

\begin{abstract} 
        We present the first view of the magnetic field structure in
the OH shell of the extreme OH/IR star OH~26.5+0.6. MERLIN
interferometric observations of this object were obtained in December
1993 in full polarisation, at 1612, 1665 and 1667~MHz. The maser spots 
show a spheroidal distribution both at 1612 and 1667~MHz, while at 
1665~MHz emission from the blue-shifted maser peak is concentrated on
the stellar position, and the red-shifted peak emission exhibits a
filamentary structure oriented on a SE-NW axis. The linear polarisation
in both main lines is rather faint, ranging from 9 to 20\% at 1665~MHz 
and from 0 to 30\% at 1667~MHz. At 1612~MHz most maser spots exhibit a 
similar range of linear polarisation although those in the outermost 
parts of the envelope reach values as high as 66\%. This is particularly 
apparent in the southern part of the shell. 
The detailed distribution of the polarisation vectors could only be obtained 
at 1612~MHz. 
The polarisation vectors show a highly structured distribution indicative of 
a poloidal magnetic field inclined by 40-60$^\circ$ to the line of sight.
The velocity distribution of the maser spots with respect to the radial 
distance is well explained by an isotropic outflow at constant velocity in the 
case of a prolate shaped spheroid envelope, also tilted about 
45-65$^\circ$  to the line of sight. 
\end{abstract}


\section{Introduction}

After leaving the main sequence, low and intermediate mass stars
experience a crucial phase in their evolution toward the white dwarf stage:
the Asymptotic Giant Branch (AGB) phase. It is at the very end of this phase 
that the star will shed most of its mass through extensive mass loss
(up to a few $10^{-4} {\rm M_{\odot} yr^{-1}}$).
The exact evolutionary sequence along the AGB to this final stage 
has not yet been resolved, but OH/IR stars are thought to trace the 
period before the proto-planetary nebula stage.
At that point, the central star is completely obscured in the optical by a 
thick dust shell built up by mass loss, but the envelope structure can be 
observed through strong emission in the ground state OH maser lines and at 
infrared wavelengths. 

While AGB stars are fairly spherical objects,  
asymmetries such as elliptical shapes or bipolar outflows are commonly
observed at the planetary nebula stage (Corradi \& Schwarz 1995).

Recently, a series of papers investigated the polarimetric structure in the
intermediate and outermost parts of the circumstellar shells of evolved stars
(Bains et al. 2003, Etoka \& Diamond 2004, Vlemmings et al. 2005, 
Vlemmings \& Diamond 2006).
Although the origin and evolution of the magnetic field is not well understood
and is currently a matter of debate, (cf. Nordhaus et al. 2007 and
reference within), this series of papers has shown the importance of the
magnetic field in shaping the circumstellar material. \\

OH~26.5+0.6 (AFGL~2205; IRAS~18348$-$0526) is an extreme OH/IR star 
at a distance of 1.37$\pm$0.30~kpc (van~Langevelde et al. 1990). Its current 
mass-loss rate has been estimated to be on the order of 
$5 \times 10^{-4}$~M$_{\odot} {\rm yr^{-1}}$ (Justtanont et al. 1996).
It has been classified as a Very-Long Period Variable OH/IR star with a 
period of 1570~days (le~Bertre 1993). 
Prior to that work, OH~26.5+0.6 has been imaged several times with the
VLA at 1612~MHz with increasing sensitivity (Baud 1981; Bowers et
al. 1983; Herman et al. 1985 and Bowers \& Johnston 1990) where a
complete ring-like structure is seen at virtually all velocities. It
has also been imaged with MERLIN (Diamond et al. 1985), where the
clumpiness of the shell was clearly revealed.

        The work presented here is part~II of a series of papers intending 
to unravel the magnetic structure around extreme OH/IR stars through 
observations in the ground state OH maser lines at 18~cm. 
The first paper of the series, Etoka \& Diamond 
(2004, hereafter paper~I), presents the magnetic field structure of the red 
supergiant NML~Cyg at 1612 and 1667~MHz. This first work has shown that a 
structured polarisation distribution exists for both lines linked with the 
geometry of the shell itself. This can be explained if the principal
driver for the shaping of the shell is the magnetic field. 

     The details of the observations and data reduction process are given in
Section~2. An analysis of the data is presented in Section~3. In Section~4 
discussion and interpretation of the results is given, while conclusions are
drawn in Section~5. 

\section{Observations \& Data reduction} 

The observations were performed on the 12$^{th}$ December 1993 at 1612, 1665 
and 1667~MHz using the 8 telescopes of MERLIN available at that time (namely
Defford, Cambridge, Knockin, Wardle, Darnhall, MK2, Lovell \& Tabley)
giving a maximum baseline of 217~km and a re\-so\-lution of
0.17~arcsec. The observations lasted 12 hours from which three hours were 
spent on calibrator sources. Data were taken in full polarisation 
mode in order to retrieve the four Stokes para\-meters.   
A bandwith of 0.5~MHz was recorded and divided into 512 channels at 
correlation, leading to a channel separation of 1~kHz, giving a velocity 
resolution of 0.18~km~s$^{-1}$. The observing programme switched at intervals 
of a few minutes between the three maser lines. The continuum source 3C84
was used to derive corrections for instrumental gain variations across 
the bandpass. 3C286 was also observed in order to retrieve the absolute 
polarisation position angles and to provide the flux density reference.
 The data reduction followed the procedure explained in paper~I, section~2.2.
        All the velocities given in this article are relative to the
local standard of rest (LSR).
\section{Analysis}

\subsection{MERLIN spectra}
Andersson et al. (1974) originally discovered the intense OH maser signal 
at 18~cm emitted by OH~26.5+0.6. This Type~II OH/IR supergiant has a 
maximum intensity observed in the 1612~MHz satellite line which is about 50 
times greater than that in the 1665/1667~MHz mainlines. \\

\begin{figure}
      \epsfig{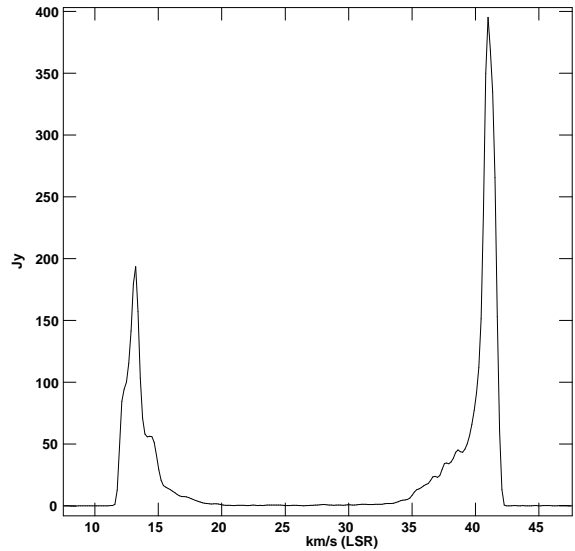}
\caption{1612~MHz spectrum of OH~26.5+0.6 in Stokes~$I$ constructed from the 
        final image.}
\label{fig: 1612 spectrum}
\end{figure}

The 1612~MHz spectrum of OH~26.5+0.6 in Stokes~$I$, constructed from the final 
image, is shown in Fig.\ref{fig: 1612 spectrum}. The spectral profile and 
the peak intensity ratio between the red- and the blue-shifted peaks of
$I_{\rm red}/I_{\rm blue}$=2 has not changed since the detection of the source 
in 1973 by Andersson et al. (1974). The intensity and the profile  
retrieved from the final map show that we recovered most of the signal. 
A faint inter-peak emission can be observed in the spectrum presented by 
Andersson et al. in the velocity range [17-22]~km~s$^{-1}$ and 
[33-35]~km~s$^{-1}$, not picked up by MERLIN.
But the general agreement in the profile and peak flux intensity between 
the two sets of data indicates that the fraction of emission potentially lost 
in extended structures is minimal. \\

\begin{figure}
      \epsfig{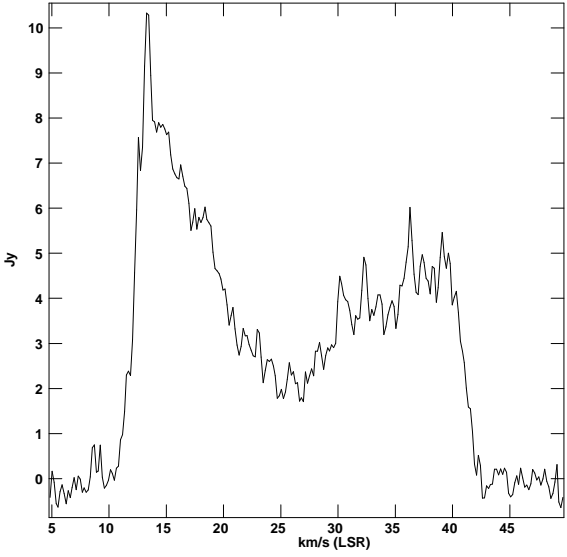}
\caption{Same as Fig~\ref{fig: 1612 spectrum} at 1667~MHz.}
\label{fig: 1667 spectrum}
\end{figure}

\begin{figure}
      \epsfig{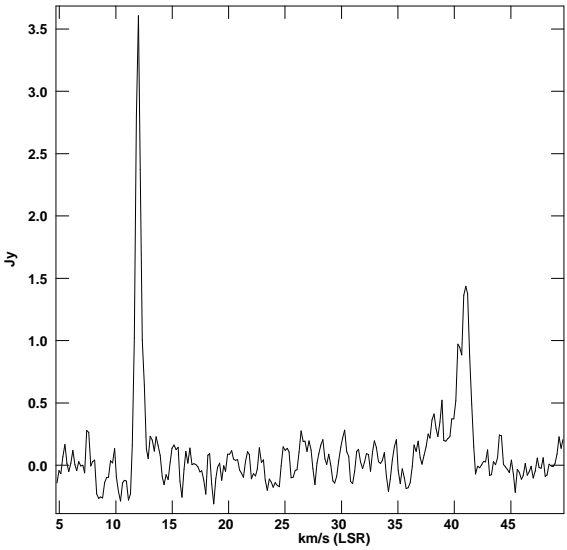}
\caption{Same as Fig~\ref{fig: 1612 spectrum} at 1665~MHz.}
\label{fig: 1665 spectrum}
\end{figure}

The spectra in Stokes~$I$, constructed from the final i\-ma\-ge at 1667 and 
1665~MHz, are shown in 
Figs.~\ref{fig: 1667 spectrum}~and~\ref{fig: 1665 spectrum} 
respectively. The profile and the peak intensity ratio between the red- 
and the blue-shifted peaks $I_{\rm red}/I_{\rm blue} \simeq$0.5 has not 
changed for the two mainlines since they were first detected. Nevertheless, we 
observed a stronger intensity corresponding to an increase of 40\% at 1665~MHz 
and 30\% at 1667~MHz from that recorded by Andersson et al. (1974) twenty 
years earlier, likely due to variability of the presumed unsaturated maser 
emission.
This assumption is strengthened by single-dish observations 
presented by Etoka \& Le~Squeren (2004) taken with the {Nan\c {c}ay} radio
telescope only three months after these MERLIN observations.
With a periodicity of nearly 1600~days this corresponds to a phase difference
of just 6\%. The spectrum profile observed at 1667~MHz by MERLIN is entirely
consistent with the single-dish observation. The spectrum observed at 1665~MHz
with the {Nan\c {c}ay} radio telescope suggests faint inter-peak emission
which is not picked up by MERLIN that could be the signature of faint
extended emission. 
But, the peak flux in both mainlines is higher in the MERLIN spectra than
in the single-dish observations (by about 10\% at 1667~MHz and 35\% at
1665~MHz). Such behaviour would be expected if both sets of observations were
taken after the OH maximum, the steeper decrease of the 1665~MHz emission 
implying less saturated emission than that at 1667~MHz.

\subsection{Maser emission extent and spot distributions}
{\sc CLEAN}ed maps of all the channels were created using the {\sc AIPS} task 
{\sc IMAGR} with a restoring beam of 0.342$\times$ 0.283~arcsec$^2$ at 1612~MHz
and 0.350$\times$ 0.140~arcsec$^2$ at 1665 and 1667~MHz.
The typical rms in Stokes~$I$ images, calculated over areas free of 
emission was about 8~mJy~beam$^{-1}$ increasing by up to 10 times that 
value for the channels with the strongest intensity. 

       The {\sc AIPS} task {\sc SAD} was used to identify maser components in 
the individual channel maps, as explained in paper~I, section~3.2.1.
At 1665~MHz, the simpli\-city of the maps was such that a 3$\sigma$ threshold 
was taken to retrieve the maser components. At 1612 and 1667~MHz, the maps 
being complex, a more stringent selection was applied for retrieving the 
components. 
A component has been accepted only if its flux density was greater than 
4$\times$rms noise of a given channel (or greater than 10$\times$rms noise in 
very complex regions).  
Similarly to paper~I, the components were then grouped into maser spots 
if they existed in more than three consecutive channels and with positional 
offsets of less than 100 mas.
With the given selection criteria, 10 maser spots were identified at
1665~MHz, 81 at 1667~MHz and 277 at 1612~MHz. 

Tables~\ref{Table: 1612 maser spot} to~\ref{Table: 1665 maser spot} present 
the flux densities  in Stokes parameters and polarisation properties of the 
maser spots fitted at 1612, 1667 and 1665~MHz respectively. 
The meaning of the 13 columns of these tables is as 
follows: column~1 gives the maser spot number. The maser spots have been 
numbered in a decreasing velocity order. Column~2 gives the peak LSR
velocity of the maser spot. Columns~3 to 6 present the corresponding I, Q, U 
and V flux densities. Column~7 presents the associated linear polarisation 
flux density. Columns~8 and 9 give the RA and DEC offsets from the pointing 
position. 
Columns~10 to 12 give the percentage of circular, linear and total 
polarisation, and finally column~13 gives the angle of the polarisation vector 
associated with the maser spot when relevant (i.e., for P$ \ge 3 \sigma$).

The strong difference between the number of maser spots found at 1612
and 1667~MHz (ratio of 4:1) is partly due to the 4$\sigma$ cutoff. This
eliminated more maser spots at 1667~MHz than it did at 1612~MHz
because most of the components at 1667~MHz are faint and did not
meet the criterion for 3 consecutive channels. 
This has an impact upon the inferred total extent of the shell at
1667~MHz, where the maser spot distribution modelling (cf. 
section~\ref{subsection: Velocity distribution}) points to a substantially 
smaller radius than that suggested by the velocity integrated image in 
Stokes~$I$.

\subsubsection{Maser emission extent} \label{subsection: maser extent}

\begin{figure}
      \epsfig{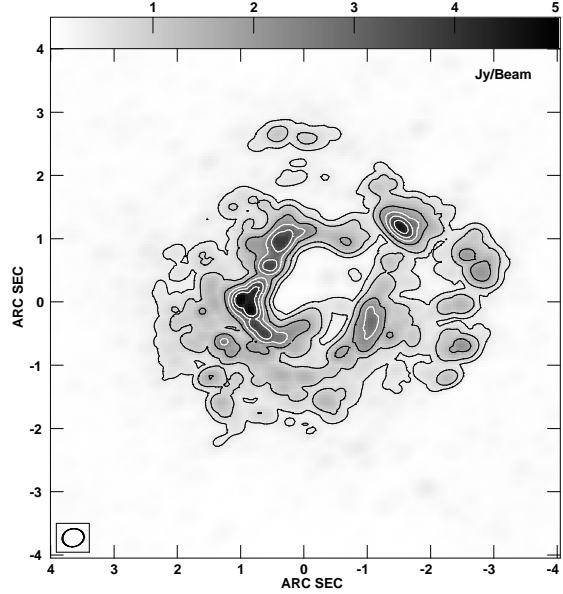}
\caption{1612~MHz velocity-integrated image of OH~26.5+0.6 in Stokes~$I$.
         The contour levels shown are 1, 2, 4, 6, 8 and 10 times 0.40
         Jy/Beam. In this velocity-integrated image the strong blue and 
         red-peak contributions (corresponding to the velocity range 
         [11:14]~km~s$^{-1}$ and [37:42]~km~s$^{-1}$) have been cut off for 
         dynamic range purposes.}
\label{fig: 1612 squash}
\end{figure}

\begin{figure}
      \epsfig{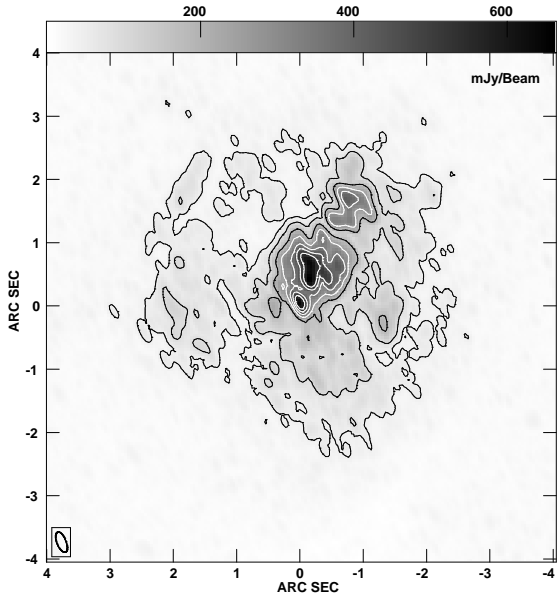}
\caption{Same as Fig.~\ref{fig: 1612 squash} for 1667~MHz. 
         The contour levels shown are 1, 2, 4, 6, 8, 10 and 12 times 
         0.044~Jy/Beam.
         This velocity-integrated image takes into account all the velocity 
         channels in which emission was detected i.e., for the velocity range 
         [11:42]~km~s$^{-1}$.}
\label{fig: 1667 squash}
\end{figure}

\begin{figure}
      \epsfig{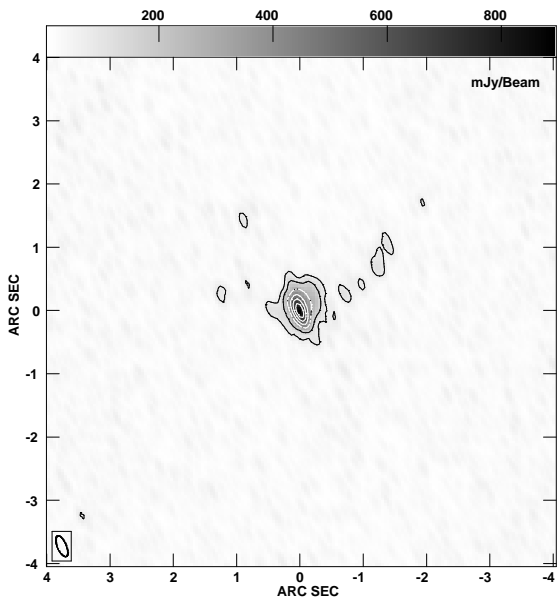}
\caption{Same as Fig.~\ref{fig: 1612 squash} for 1665~MHz. 
        The contour levels shown are 1, 2, 4, 6, 8 and 10 
        times 0.075~Jy/Beam. This velocity-integrated image takes into account 
        all the velocity channels in which emission was detected i.e., for the 
        velocity range [11.5:41.5]~km~s$^{-1}$.}
\label{fig: 1665 squash}
\end{figure}

The velocity-integrated images for the 1612, 1667 and 1665~MHz maser emission 
are presented in Figs.~\ref{fig: 1612 squash},
\ref{fig: 1667 squash} and \ref{fig: 1665 squash} respectively.
In Fig.~\ref{fig: 1612 squash}, the strong blue- and red-peak contributions 
(corresponding to the velocity range [11:14]~km~s$^{-1}$ and 
[37:42]~km~s$^{-1}$) have  been cut off for dynamic range purposes.
The 1612 and 1667~MHz extents are about 5~arcsec which corresponds to a linear 
extent of $\sim$7000~AU at 1.37~kpc. 
At 1612~MHz, the bulk of the emission describes a ring centred about 
+0.5~arcsec in $\delta$ from the optical stellar position.
The 1665~MHz central core emission lies within an area less than 1.5~arcsec 
across.
Including the very faint maser spots observed East and West, the total extent 
of the 1665-MHz emission is still less than 4~arcsec.

\begin{figure*}
      \epsfig{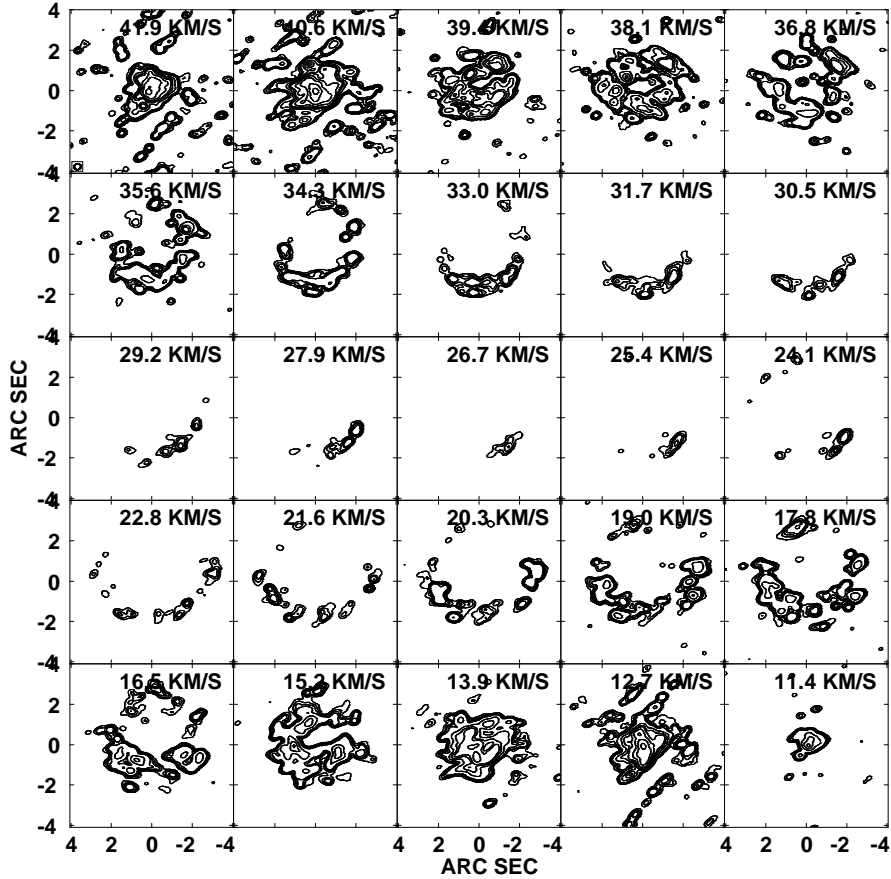}
\caption{1612~MHz maps of OH~26.5+0.6 in Stokes I. Each map is an integration 
         of 7 channels (i.e., leading to a map separation of 1.27~km~s$^{-1}$). 
         The contour levels shown are 3, 4, 5, 7, 10, 30, 60, 90, 180,
         360, 720 and 1440 times 0.019 Jy/Beam. The choice of the contours
         has been made so that the relatively faint emission can be seen in the
         velocity range [30-20]~km~s$^{-1}$. But this also stresses the 
         dynamic range problem in channels where strong emission is present, 
         that occurs around 41~km~s$^{-1}$ particularly but also around 
         12~km~s$^{-1}$ to a lesser extent.
}
\label{fig: 1612 maps}
\end{figure*}

\begin{figure*}
      \epsfig{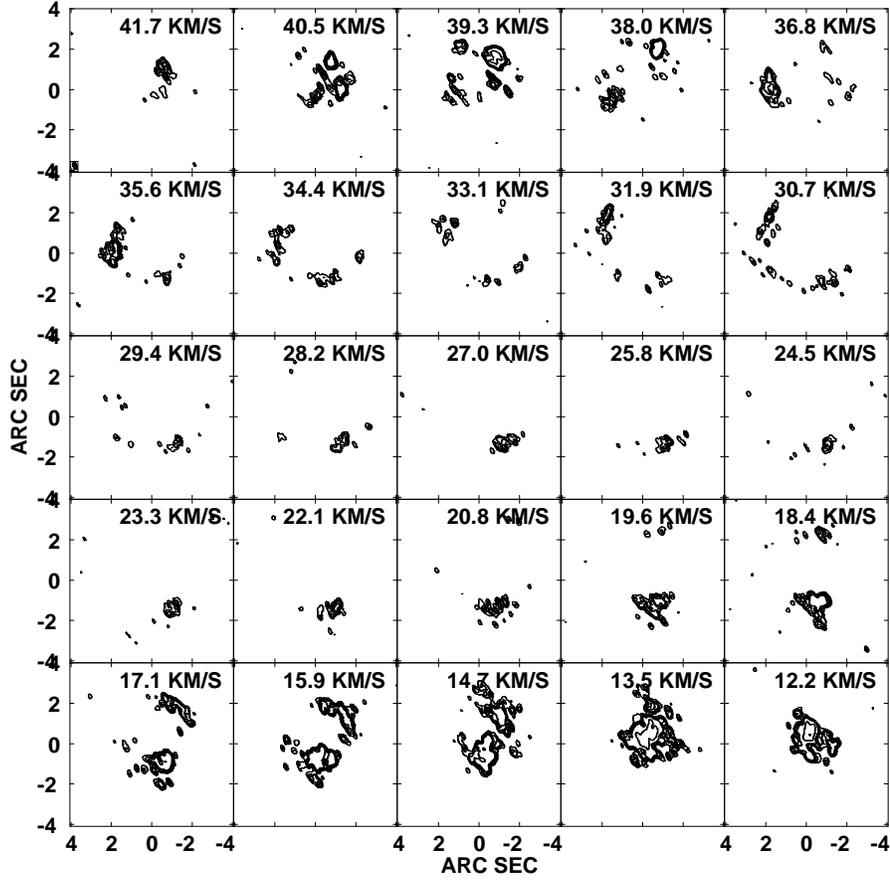}
\caption{Same as Fig.\ref{fig: 1612 maps} for the 1667~MHz emission.
         The contour levels shown are 3, 4, 5, 7, 10 and 30 times 
         0.0042 Jy/Beam.}
\label{fig: 1667 maps}
\end{figure*}

Figures~\ref{fig: 1612 maps}~and~\ref{fig: 1667 maps} show the maps of the 
distribution of the emission integrated over a velocity interval of 1.27
and 1.23~km~s$^{-1}$ at 1612 and 1667~MHz respectively.  
From these figures a certain number of physical properties concerning
the geometry and the dynamic of the shell can be inferred:
\begin{itemize} 
  \item the ellipsoidal nature of the shell is revealed with an axis ratio 
        of $\sim$0.80 and a projected major axis position angle of 
        $20^{\circ} \pm 5^{\circ}$. It is clearly apparent at 1612~MHz in the 
        velocity range [23:18]~km~s$^{-1}$. 
  \item at 1612~MHz, the maser emission distribution along the velocity
        channels is consistent with a radially expanding shell;
  \item at 1667~MHz, there is a hint of a deviation from the
        uniform radial expansion in the red-shifted emission since
        the `central spot' expected at V=+41.7~km~s$^{-1}$ is not observed. 
        Instead, a maser spot approximately 1~arcsec \-off-centre is observed;
  \item a clear asymmetry is observed, as an
        incomplete ring structure can be seen both at 1612 and
        1667~MHz. At 1612~MHz, there is no detectable maser 
        emission radiating from the north of the shell
        in the velocity range [33:20]~km~s$^{-1}$, while at 1667~MHz 
        virtually no emission is observed in both the NW and SE
        quadrants in the same velocity range.
\end{itemize}

\subsubsection{Location of the star} \label{subsection: star location}
At the time of the observations, MERLIN data were not routinely
phase-referenced. Therefore, an assumption regarding the location of the 
central star has to be made.  
As mentioned in paper~I, amplification of the stellar radiation by the bluest
emission of the spectrum has been observed for various types of OH/IR
emitters, strongly suggesting that this feature marks the location of the star
(Norris et al. 1984; Sivagnanam et al. 1990; van~Langevelde et al. 2000).  
Therefore, the maser component belonging to the blue-shifted peak and located 
at the centre of the maser distributions at 1612 and 1667~MHz, at a velocity of 
11.4~km~s$^{-1}$ and 12.2~km~s$^{-1}$ respectively 
(cf. Figs.~\ref{fig: 1612 maps} \&~\ref{fig: 1667 maps}) is quite likely to 
be over the stellar position. Similarly, the blue-shifted peak at 
V=12~km~s$^{-1}$ at 1665~MHz is taken to be centred at the stellar position.
 
This assumption has been followed for the analysis of the data presented in 
this article.
 
\subsubsection{Maser spot distributions}

\begin{figure}
      \epsfig{file=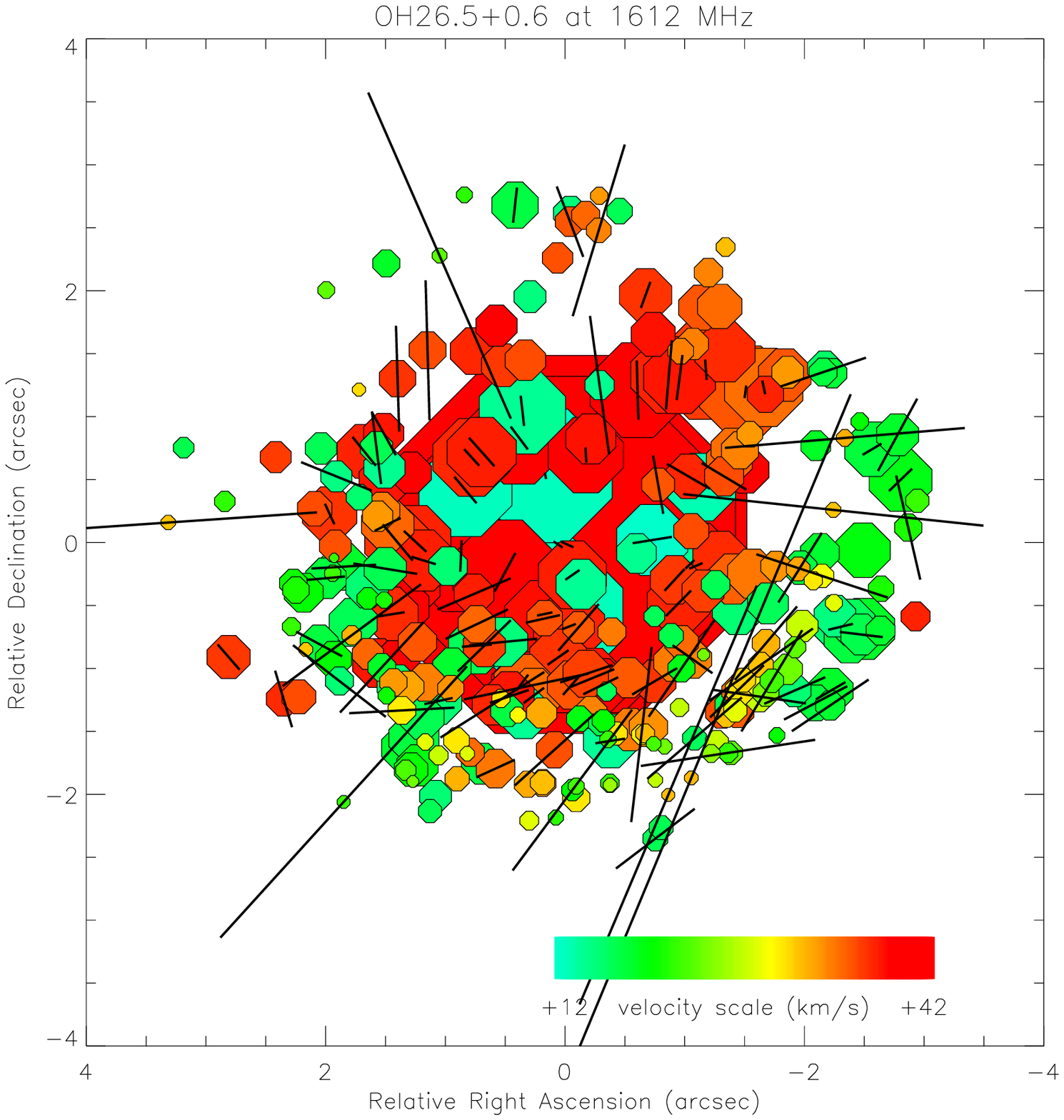,width=7.65cm,angle=0}
\caption{1612~MHz maser spot distribution of OH~26.5+0.6. The area of a
symbol is proportional to the maser spot intensity. The colour scale
indicates velocity. Also plotted are the polarisation vectors
associated with each maser spot with P$>3\sigma$. 
The length of a vector is proportional to the percentage of linear 
polarisation.}
\label{fig: 1612 maser spot}
\end{figure}

\begin{figure}
      \epsfig{file=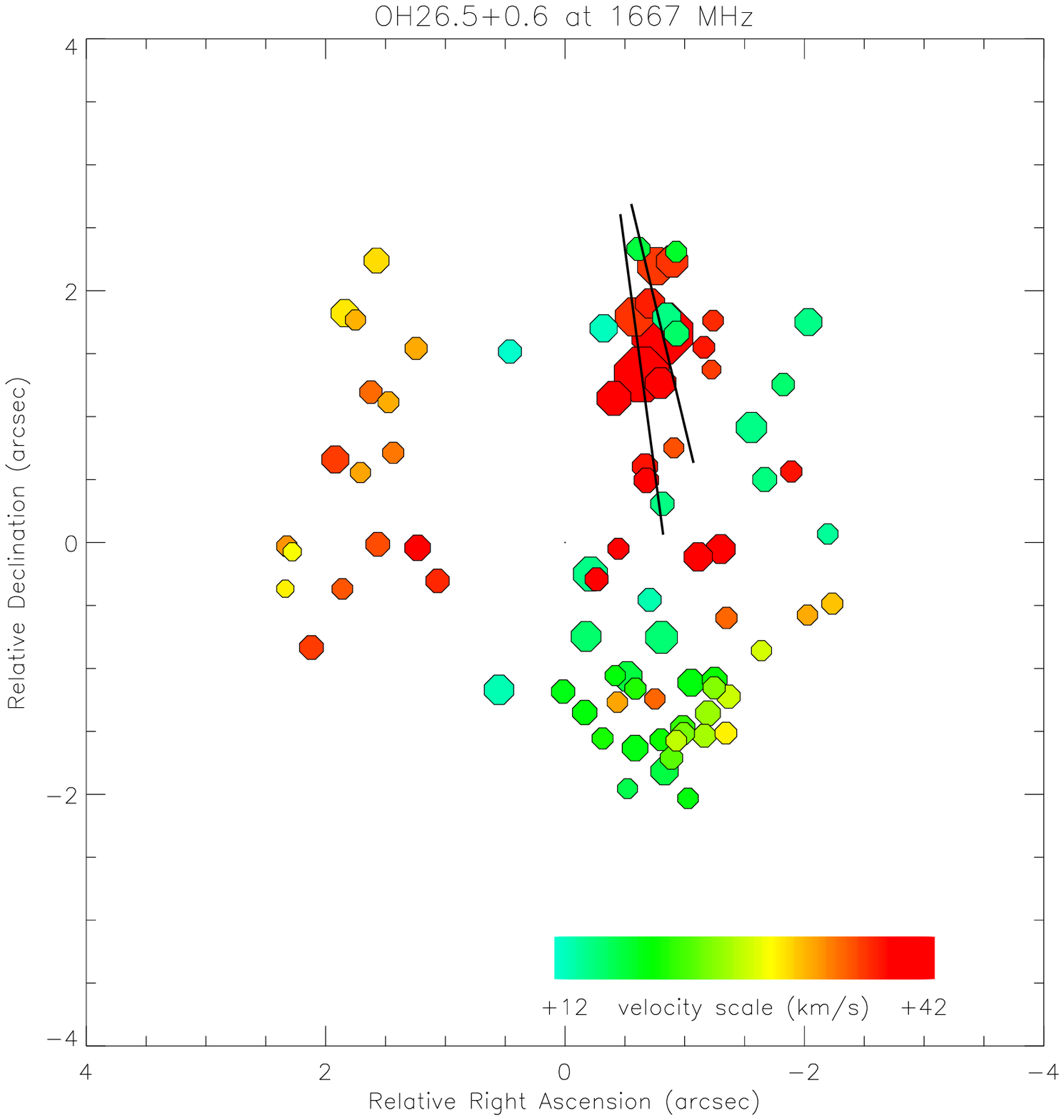,width=7.65cm,angle=0}
\caption{Same as Fig.~\ref{fig: 1612 maser spot} for the 1667~MHz.}
\label{fig: 1667 maser spot}
\end{figure}

\begin{figure}
      \epsfig{file=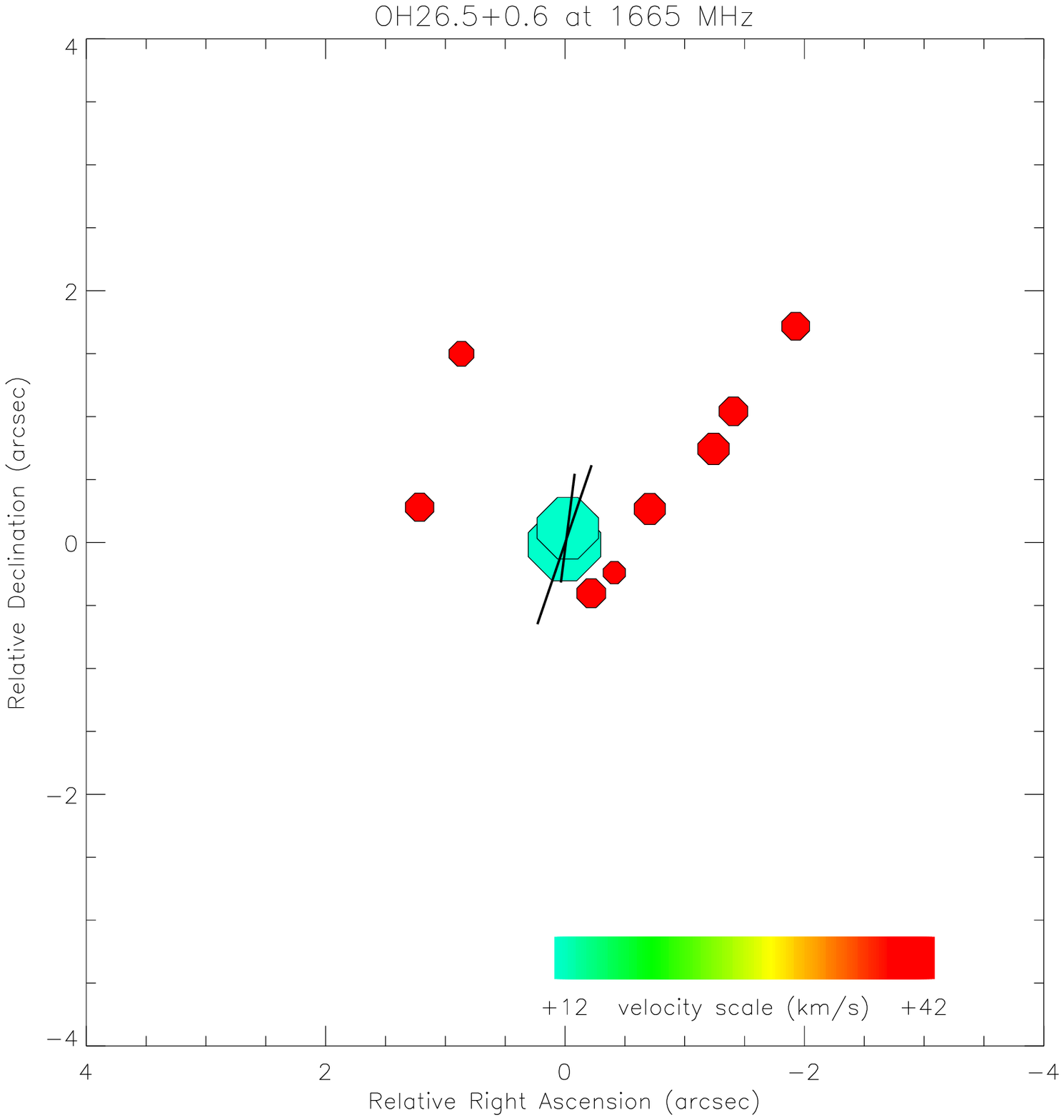,width=7.65cm,angle=0}
\caption{Same as Fig.~\ref{fig: 1612 maser spot} for the 1665~MHz.}
\label{fig: 1665 maser spot}
\end{figure}

Figures~\ref{fig: 1612 maser spot}~to~\ref{fig: 1665 maser spot} present the 
maser spot distributions observed at 1612, 1667 and 1665~MHz respectively.
The maser spots show a spheroidal distribution both at 1612 and 1667~MHz. 
But, 1612~MHz shows by far the most complex spatial velocity distribution, 
in which the outermost part of the distribution is dominated by blue-shifted 
maser spots. 

        At 1667~MHz, there is a rough gradient in the velocity distribution 
of the maser spots, such that the red-shifted spots are found in the N-NW part 
of the shell while the blue-shifted masers are found in the centre and S-SE 
part of the shell, contrasting with the 1612~MHz structure.

At 1665~MHz, emission from the blue peak is concentrated on the stellar 
position and the red-peak emission exhibits a filamentary structure oriented 
on a SE-NW axis. 

\subsection{Polarimetry}\label{Pol section}

\begin{figure}
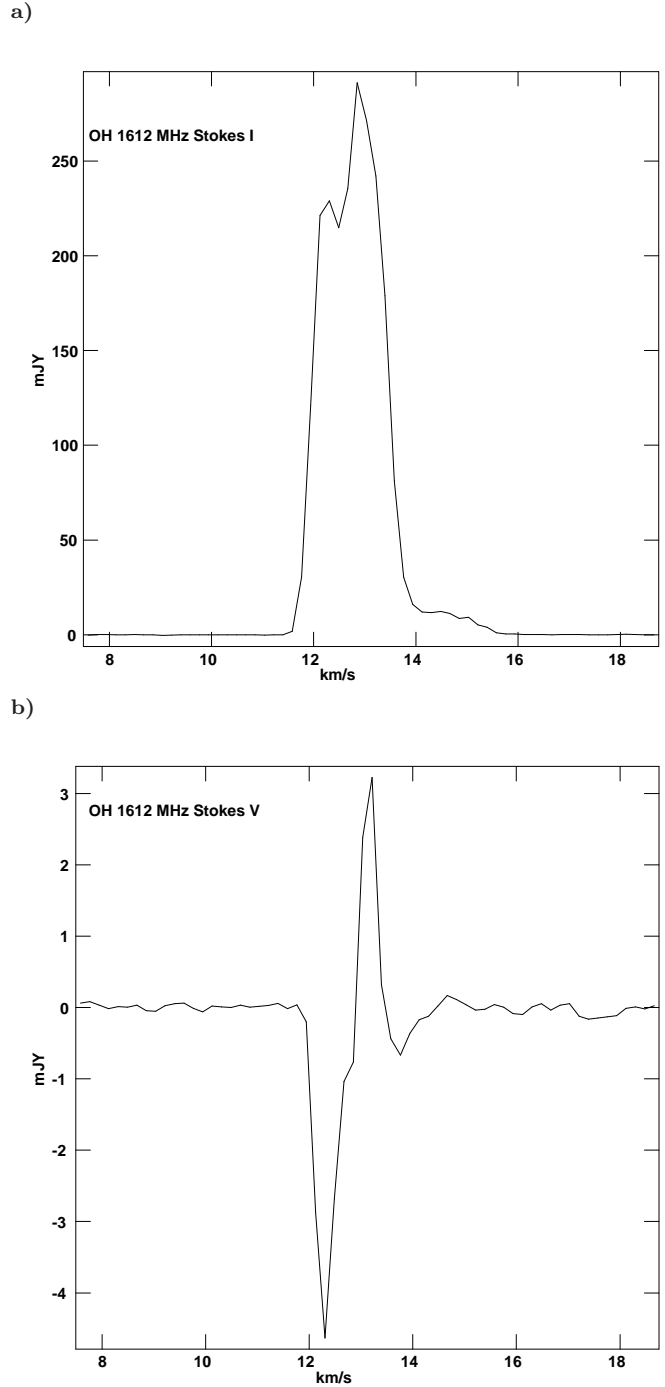

\noindent {\bf a)} \\
\vspace*{-0.5cm} \epsfig{file=SEtoka_fig12a.ps,width=8.75cm,angle=0}   
\noindent {\bf b)} \\
\vspace*{-0.5cm} \epsfig{file=SEtoka_fig12b.ps,width=8.75cm,angle=0}
   \caption{Stokes~$I$ and $V$ spectrum for a Zeeman pair of a 
     blue-shifted component at the relative position $\delta$RA=-90~mas and 
     $\delta$Dec=+180~mas. The separation between the 2 pairs provides an 
     estimate of the magnetic field $B =-3.7\pm 0.3$~mG that is pointing 
     towards the observer.}
     \label{fig: StokesV 1612 MHz}
   \end{figure}

A few possible Zeeman patterns were found 
(cf. Fig.~\ref{fig: StokesV 1612 MHz}) leading to 
a magnetic field at the location of the OH shell $B =-3.7\pm 0.3$~mG. 
At a similar distance, the magnetic field strength in NML~Cyg has been 
estimated to be also about 3~mG (paper~I). \\

The linear polarisation in both mainlines is rather faint, ranging from 
9 to 20\% at 1665~MHz and from 0 to 30\% at 1667~MHz. 
At 1612~MHz, most maser spots exhibit a similar range 
but the outermost maser spots tend to exhibit a greater degree of linear 
polarisation, reaching values as high as 66\%. 
The strongest linearly polarised components belong to the southern
part of the shell. \\

\begin{figure}
\epsfig{file=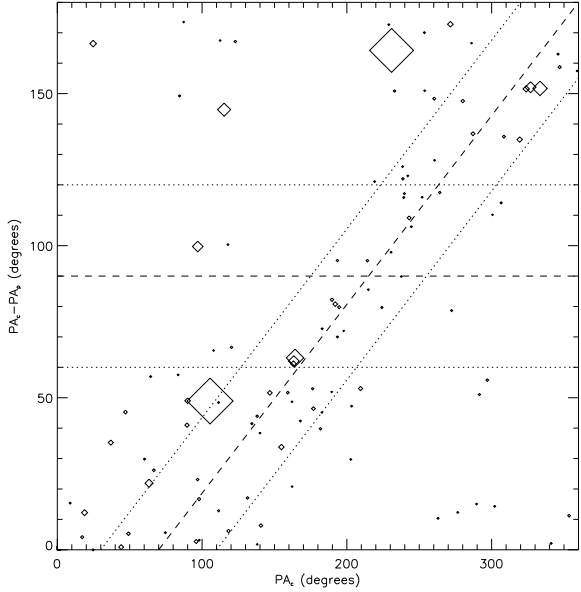,width=8.25cm,angle=0}
\caption{Difference in angle between the polarisation vector orientation 
and the radial direction (given by PA$_c$~-~PA$_p$) versus the position angle
of the maser components (PA$_c$) at 1612~MHz. The size of the symbol is 
proportional to the corresponding maser spot intensity. The dashed line 
represents the best fit for the general trend observed. The dotted lines 
delineate a deviation of 20\% from the best fit.}
\label{fig: PA vs RA}
\end{figure}

        The information concerning the magnetic field structure at the 
location of the OH maser emission is displayed in 
Figs.~\ref{fig: 1612 maser spot}~to~\ref{fig: 1665 maser spot} via the 
vectors of polarisation which reveal the electric field plane at the polarised
radiation. 
        At 1667~MHz and 1665~MHz, only two maser spots had
polarised flux P$\ge 3\sigma$. 
At 1612~MHz, out of the 277 maser spots detected, 106 had P$>3\sigma$. 
Therefore the detailed polarisation vector distribution could only be 
obtained for the latter transition. 
The vector distribution reveals a highly ordered polarisation field.
Overall, the polarisation vectors show a mixture of radial and tangential 
distributions: the polarisation vectors in the N-NE part of the shell are 
radial while those in the S-SW are generally tangential. The position
angle (PA) of the projected axis along which the tangential/radial
separation occurs is about PA=100$^{\circ} \pm 10^{\circ}$.
This is illustrated by Fig.~\ref{fig: PA vs RA} which presents a view
on how the polarisation angles are related to the radial
direction (PA$_{\rm c}$) of their associated maser spot at 1612~MHz. 
More precisely, this figure shows the deviation of the vectors of
polarisation from the tangent. The general trend observed clearly
shows the change in direction of the polarisation vectors with orientation. 
A similar dichotomy in the orientation of the polarisation vector was
observed by Boboltz (1997) for the Mira star R~Aqr in SiO.

Goldreich et al. (1973) showed that in the limiting case of strong 
saturated maser emission and overlapping of the Zeeman components, a flip of 
90$^\circ$ in the plane of polarisation occurs when the angle between 
the magnetic field direction and the line of sight
is close to the critical angle of $\sim$55$^\circ$. 
Following Elitzur (1996) we can estimate the ratio $\chi_{B}$ for the 
significance of the Zeeman splitting: \\

\begin{equation}
\chi_{B}=\frac{\Delta \nu_{B}}{\Delta \nu_{D}}=14g\lambda \frac{B}{\Delta v_{D}}
\label{Equation: ratio of Zeeman to Doppler}
\end{equation}
where the Lande factor $g$=0.935 for the $^2 \Pi_{3/2} \, J=3/2$ (ground state) 
transitions of OH.

For a magnetic field strength of 3.7~mG, 
$\chi_{B}=\frac{0.9}{\Delta v_{D}} < 1$ if 
$\Delta v_{D} > 0.9$~km~s$^{-1}$, which has to be compared with the width of 
the line, found to be about 2~km~s$^{-1}$ 
(cf. Fig.~\ref{fig: StokesV 1612 MHz}).
A magnetic field of the order of a few mG in the case of maser emission in 
the ground state of OH implies a Zeeman splitting exceeding the stimulated rate 
(i.e., g$\Omega>$ R). In addition, saturation of the 1612~MHz line implies a 
stimu\-lated emission rate exceeding the decay constant (i.e, R$>\Gamma$).
We therefore interpret the flip of the plane of linear polarisation observed 
in maser spots at 1612~MHz as due to the magnetic field being inclined by 
an angle close to $\theta_{crit} \sim 55^\circ$ to the line of sight.
 Such a configuration accounts for the change of orientation of 
the polarisation vectors between tangential and radial as observed here.
Such a flip in the polarisation angle has indeed already been observed in SiO 
and H$_2$O respectively (Kemball \& Diamond 1997, Kemball et al. 2009 and 
Vlemmings \& Diamond 2006). It has never been observed in OH so far though 
since usually the Zeeman pattern is fully separated. \\

\subsection{Velocity distribution}\label{subsection: Velocity distribution}

        The $V$=f($\theta$) distributions at 1612, 1667 and 1665~MHz are 
shown in Figs.~\ref{fig: veloc vs radial dist}(a,b,c), in which the maser 
component corresponding to the blue-shifted peak for all 3 lines has been 
taken to be at the stellar position 
(cf. Section~\ref{subsection: star location}) and the stellar 
velocity is taken to be V$_{\rm star}=+27$~km~s$^{-1}$ .

\subsubsection{Comparison with the standard model}
The simple model for a uniformly expanding spherical thin shell 
(Reid et al. 1977) is given by~:
\begin{equation}
\left(\frac{\theta}{\theta_S}\right)^2 + 
        \left(\frac{V-V_{\rm star}}{V_{\rm exp}}\right)^2 =1
\label{eq: thin shell model}
\end{equation}
%

\begin{figure}
\noindent {\bf a)} \\
\vspace*{-0.4cm}    \epsfig{file=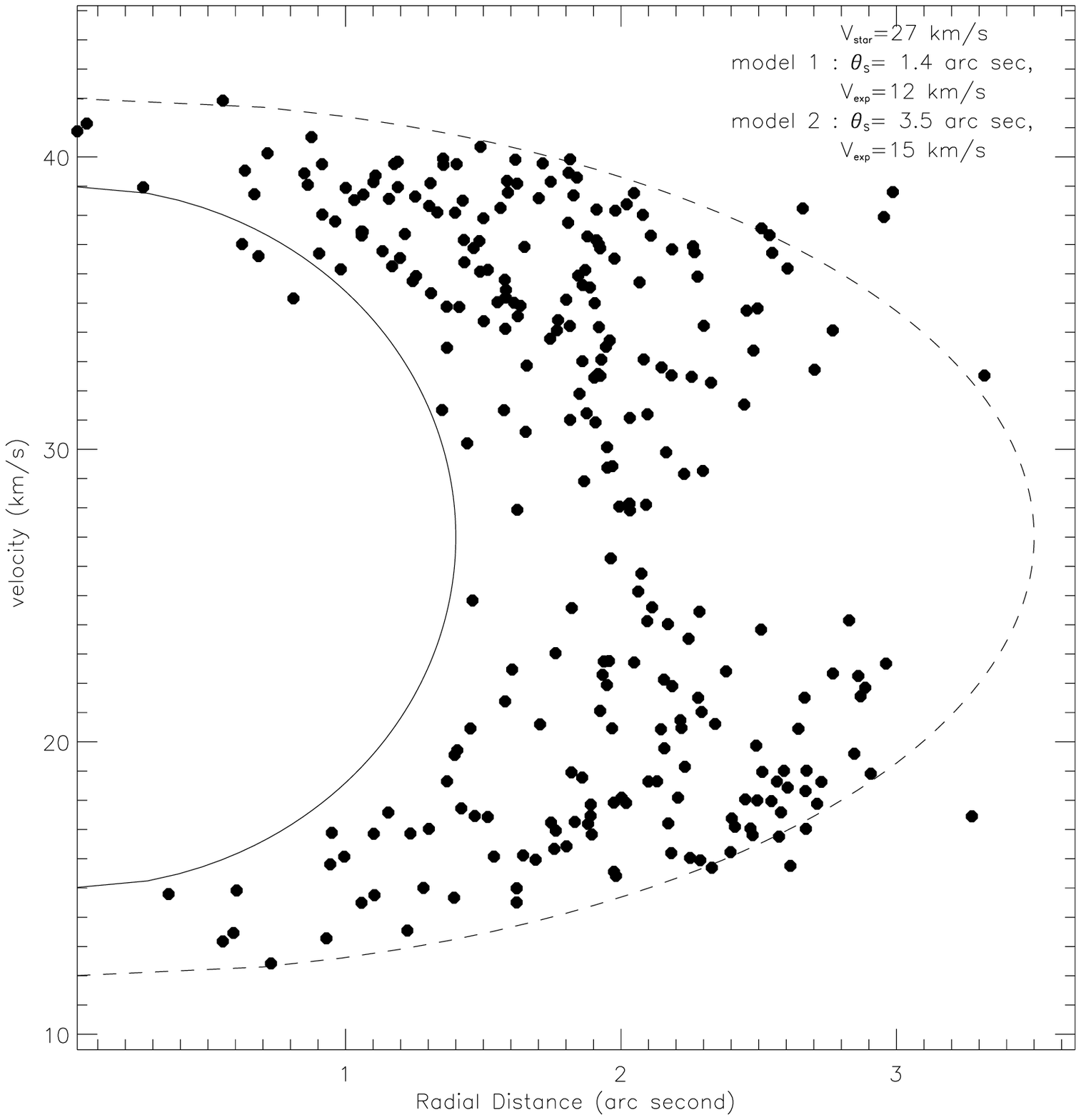,width=7.25cm,angle=0} \\
\noindent {\bf b)} \\
\vspace*{-0.4cm}    \epsfig{file=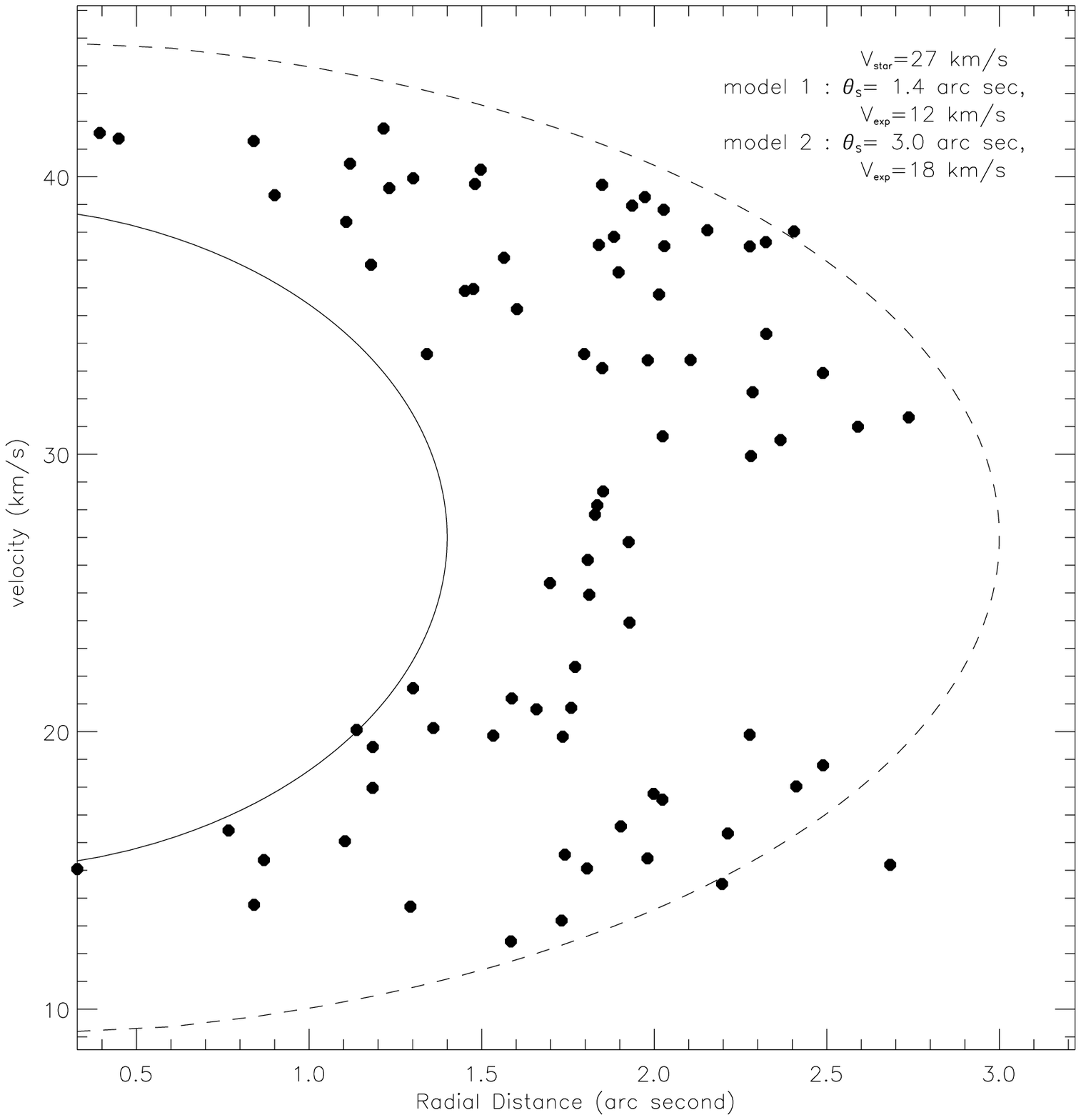,width=7.1cm,angle=0} \\
\noindent {\bf c)} \\
\vspace*{-0.4cm}   \epsfig{file=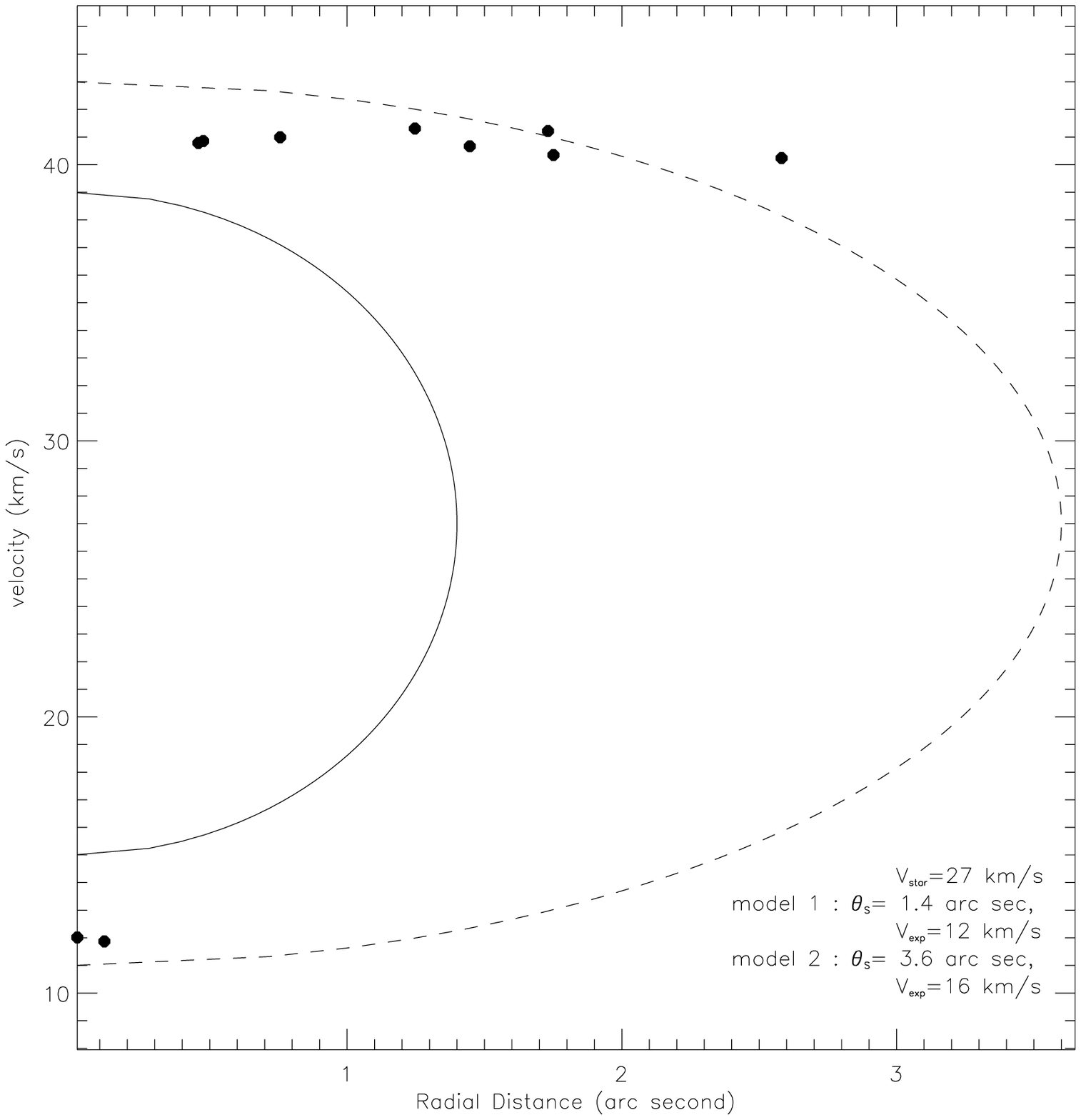,width=7.1cm,angle=0} \\
\caption{Velocity distribution of the maser spots versus the radial
distance from the star projected on the plane of the sky at {\bf a)}
1612~MHz, {\bf b)} 1667~MHz and {\bf c)} 1665~MHz. The inferred location 
of the star is RA=0~mas Dec=0~mas which is the location of the blue shifted 
peak in each line as explained Section~\ref{subsection: star location}.}
\label{fig: veloc vs radial dist}
\end{figure}

%
Where $\theta_S$ is the shell radius, $V_{\rm star}$ is the velocity of the 
star and $V_{\rm exp}$ the expansion velocity.
Generally, this model provides a good explanation for the velocity 
distribution observed in OH/IR stars (Habing 1996). On the 3 figures are 
displayed the two best fits for the lower and upper boundaries of the radial 
velocity distribution. 
These are as follows: 
\begin{itemize}
\item at 1612~MHz: $\theta_S=1.4$~arcsec and V$_{\rm exp}=12$~km~s$^{-1}$ 
      for the lower boundary (i.e., model~1 in 
      Fig.~\ref{fig: veloc vs radial dist}a) and $\theta_S=3.5$~arcsec and 
      V$_{\rm exp}=15$~km~s$^{-1}$ for the upper boundary 
      (i.e., model~2 in Fig.~\ref{fig: veloc vs radial dist}a);
\item at 1667~MHz: $\theta_S=1.4$~arcsec and V$_{\rm exp}=12$~km~s$^{-1}$ 
      for the lower boundary (i.e., model~1 in 
      Fig.~\ref{fig: veloc vs radial dist}b) and $\theta_S=3.0$~arcsec and 
      V$_{\rm exp}=18$~km~s$^{-1}$ for the upper boundary 
      (i.e., model~2 in Fig.~\ref{fig: veloc vs radial dist}b);
\item at 1665~MHz: $\theta_S=1.4$~arcsec and V$_{\rm exp}=12$~km~s$^{-1}$ 
      for the lower boundary (i.e., model~1 in 
      Fig.~\ref{fig: veloc vs radial dist}c) and $\theta_S=3.0$~arcsec and 
      V$_{\rm exp}=16$~km~s$^{-1}$ for the upper boundary 
      (i.e., model~2 in Fig.~\ref{fig: veloc vs radial dist}c).
\end{itemize}

The simple model provides a reasonable explanation for the expansion of 
the inner shell. Also, it allows us to infer that the OH maser region 
has a certain thickness, which in the context of the standard model would be 
about 1.5-2~arcsec, and that acceleration is still taking place in the 
outer part of the circumstellar envelope. Both the 1612 and 1667~MHz 
distributions lead to similar results. But clearly, this model does not 
describe well the expansion of the outermost maser components. Indeed, the 
maximum shell radius is not observed at the stellar velocity, 
V$_{\rm star}=+27$~km~s$^{-1}$, but at two values equidistant from the 
stellar velocity, on either side of it. 
Sensitivity is not the cause since maser spots were indeed found
around the stellar velocity range both at 1612 and 1667~MHz
(cf. Table~\ref{Table: 1612 maser spot} and 
Table~\ref{Table: 1667 maser spot}).  \\

\subsubsection{Comparison with the models of Bowers (1991)}

Bowers (1991) produced a series of kinematic models as a tool 
to analyse complex aspherical outflows observed in the 
cirmcumstellar shell of evolved stars. In these models, the maser 
emission is uniformly distributed throughout ellipsoidal shells 
with various orientations to the line of sight. The effect of 
rotation and radial acceleration that might be present in the 
velocity field is also taken into consideration. These models 
produce a large variety of possible $\theta(V)$ and $I(V)$ 
curves that may be applicable to stellar outflow. As an 
application of these models, the author successfully describes 
the shell structure of 3 different types of aspherical outflows 
commonly observed at the late stages of stellar evolution.

\begin{figure}
 \hspace*{-0.75cm}    \epsfig{file=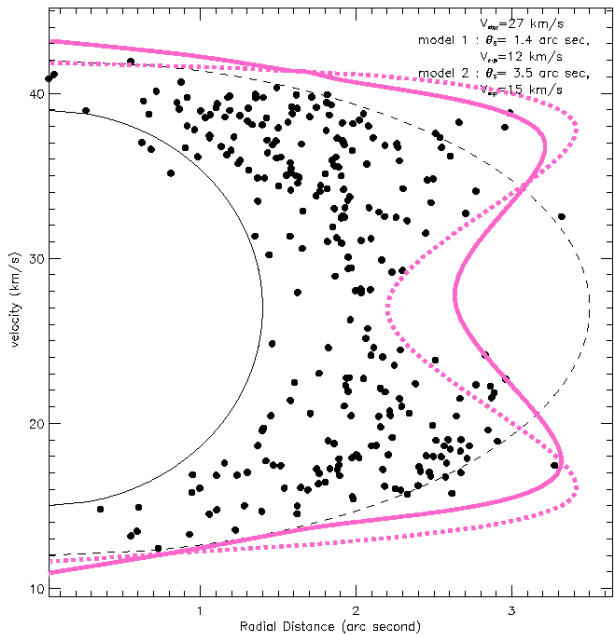,width=10cm,angle=0}
     \caption{The radial velocity distributions at 1612~MHz on which is
      overlaid on top of the standard models (thin black lines), a series 
      of schematic models in magenta illustrative of an isotropic 
      outflow at a constant velocity in the prolate case, where the 
      inclination of the spheroid from the line of sight is $i=45^{\circ}$ 
      (thick dotted line) and  $i=65^{\circ}$ (thick continuous line).
}
     \label{fig: vel dist and models}
   \end{figure}

 Overlaid on the 1612~MHz velocity distribution reproduced in 
Fig.~\ref{fig: vel dist and models}, are two 
schematic models consistent with the distribution observed. The dotted 
line presents a schematic illustration of Bowers' model for an isotropic 
outflow at a constant velocity in the prolate case, where the inclination of 
the spheroid from the line of sight is $i=45^{\circ}$, while the continuous 
line is for the spheroid tilted from the line of sight by $i=65^{\circ}$. In 
the $i=65^{\circ}$ case, a standard double-peak spectral profile is expected 
while for $i=45^{\circ}$, for both the red and blue peaks, a double-component 
structure is expected, with the component closer to the stellar velocity 
being fainter than the external one. The intensity of the internal
components increases when $i$ decreases. There is indication of such an internal
component structure in the 1612~MHz spectrum shown in 
Fig.~\ref{fig: 1612 spectrum}.
Consequently, both the $V$=f($\theta$) distribution and the spectral profile 
$I$=f($V$) are well explained by an isotropic outflow at constant 
velocity in the ellipsoidal prolate case, with an inclination to the 
line of sight between {45\degr} and {65\degr}. 
Note that none of the other cases in the series of kinematic models 
presented by the Bowers are able to explain simultaneously the 
$V$=f($\theta$) and the spectral profile $I$=f($V$) we observe.

\section{Discussion}

\subsection{Actual stage of evolution of OH~26.5+0.6}

The work of Sevenster (2002) and  Ortiz et al. (2005) based on the 
MSX catalogue at the mid-infrared (MIR) wavelengths 
8.3, 12.1, 14.7 and 21.3~$\mu$m showed that [8.3-14.7] vs [14.7-21.3] is 
better suited than the IRAS colour-colour diagram to separate AGB from 
post-AGB stars and that [15-21] vs [8-12] is the most efficient index to 
separate the four main classes of OH/IR objects: PPNs, SFRs, AGB and post-AGB
stars. In the light of these results, we calculated the [8.3-12.1], [8.3-14.7] 
and [14.7-21.3] colour indices for OH~26.5+0.6 from the MSX measurement for 
this source to be 0.5669, 1.0832 and 0.0624 respectively. We then used these 
indices to locate OH~26.5+0.6 in the colour-colour diagrams of 
Ortiz et al. (2005) and Sevenster (2002).

Those values put OH~26.5+0.6 in the bluer OH/IR group of Ortiz et al. (2005),
according to their classification based on the [8.3-14.7] and [14.7-21.3]
indices. While it places it in the bulk of the AGB stars in Sevenster's 
(2002) diagram, based on the [14.7-21.3] and [8.23-12.13] colour-colour indices.
This indicates that OH~26.5+0.6 is definitely still on the AGB at the
present time.

Nevertheless, its IRAS [60-25] and [12-25] colour indices  
combined with its 18~cm OH maser properties attest to a thick circumstellar 
envelope characteristic of a rather evolved AGB star 
(Etoka \& Le~Squeren 2004). 
Its infrared and OH characteristics resemble those observed for red OH/IR 
supergiants. 
It is one of the brightest OH maser emitters in our Galaxy, and 
in order to account for its infrared SED, Justtanont et al. (1996) evaluated 
its Main Sequence mass to be in the order of 8~M$_{\odot}$. 
The latter authors also estimated the current mass-loss rate of  OH~26.5+0.6 
to be on the order of $5 \times 10^{-4}$ M$_{\odot}$/yr, triggered by the 
onset of a superwind phase just 150 years ago. 

All this indicates that this object is at the tip of the AGB.   
This makes OH~26.5+0.6 a particular and important object in terms of 
stellar evolution: a junction object between an intermediate- and a 
high-mass evolved object on the verge of leaving the AGB towards the 
planetary nebula phase.

\subsection{Distance considerations}

The distance of OH~26.5+0.6 was calculated by van~Langevelde et al. (1990) 
using phase lags which rely on the assumption of maser saturation, spherical 
symmetry and the thin shell model. 
The assumption of saturation for the OH 1612~MHz emission of Type~II OH/IR 
has been demonstrated
(Harvey et al. 1974, Etoka \& Le~Squeren 2000).
Nonetheless it is clear from our results that the OH shell of OH~26.5+0.6 
deviates from strictly spherical symmetry and the thin shell model condition. 
This deviation has consequently an impact on the actual distance inferred by 
van~Langevelde et al. (1990). Indeed, as stated by those authors, 
phase lag measurement relies mainly on the reddest and bluest part of the 
spectral profile while angular diameter determination from interferometric 
measurement relies on the velocity ranges closest to the stellar velocity.
While the blue and red peaks, under the hypothesis of radial expansion, trace 
the front and rear caps of the shell along the line of sight, the velocity 
ranges responsible for the emission in the plane of the sky, that is in a 
perpendicular direction, are more internal. And the difference in depth is 
likely to be more important in the case of a thick shell.
A direct consequence of the divergence from thin shell model, but still 
assuming spherical geometry, is the following: \\ 

\begin{equation}
\tau(D_{\rm phase \, lag}) \ge \tau(D_{\rm int})
\nonumber
\end{equation}
where $\tau(D_{\rm phase \, lag})$ is the time travel difference for emission 
coming from the front and rear caps of the shell  separated by
$D_{\rm phase \, lag}$ 
and, $\tau(D_{\rm int})$ is the time travel difference for 
emission coming from the total extent of the shell, $D_{\rm int}$, as obtained 
from interferometric mapping (i.e., the time that would have been inferred if 
the observer was in the plane of the sky seeing the same object at an angle of 
90$^\circ$).
But because we are diverging from sphericity, with a prolate spheroid, this 
also implies that  $D_{\rm int} > D_{\rm phase \, lag}$ which would have a 
compensating effect in this particular case.
Herman et al. (1985) consider the impact of deviation from sphericity and the 
thin shell model on distance determination with phase lag. They conclude that 
an asymmetry of $\le$~20\% and the thickness of $\le$~20\% would set a limit 
of $\sim$ 10\% in the determination of the distance of OH/IR objects subject 
to these deviations. In our case, the more marked thickness of the shell would 
quite probably reduce the accuracy down to 20\%.
This shows the importance of better constraints on the actual shell properties 
(i.e., geometry and thickness) in order to get a more accurate distance 
determination from phase lag. Nonetheless, an uncertainty of 20\%, already 
claimed by van~Langevelde, would not have a major impact on the analysis 
presented here.

\subsection{Faraday rotation}

 Faraday rotation could potentially rotate the polarisation vectors by a 
substantial angle if the radiation is propagating into an ionized medium. 
The change in the angle is given by: \\ 

\begin{equation}
\Delta \theta_{\rm Faraday}= {\rm RM} \lambda^2
\label{Equation:Faraday Angle}
\end{equation}

\noindent
where RM is the rotation measure given by:\\
\begin{equation}
 {\rm RM}= 0.812 \int\limits_{0}^{d} \lbrack \frac{n_e(s)}{{\rm cm^{-3}}} \rbrack 
   \lbrack \frac{B(s)}{\mu G} \rbrack 
   (\frac{ds}{pc})
\label{Equation:RM}
\end{equation}
(cf. Noutsos et al., 2008 their Equation~4). \\

 The two main sources of Faraday rotation which could have an impact on the 
overall polarisation vectors are foreground rotation due to propagation 
in the interstellar medium (ISM) and, more importantly, rotation within the 
shell itself due to ionized material which would affect the linear 
polarisation of the red-shifted maser spots.
 We investigated these possible causes of Faraday rotation:
\begin{itemize}
\item Noutsos (private communication) calculated the RM 
at the location of OH~26.5+0.6 from adjacent pulsars to be 
RM $\sim$ 7 rad~m$^{-2}$. Such a value would produce a 
Faraday rotation of about $\Delta \theta_{\rm Faraday}=$13$^{\circ}$.
Nonetheless, it has to be acknowledged that the uncertainty on this value 
is quite high due to the nature of the ISM and depends strongly on the 
density model adopted.In particular, adopting the electron
density model (NE2001) of Cordes \&  Lazio (2002) would lead to a
value of RM~$\sim$ 40 rad~m$^{-2}$ and potentially a rotation of
$\Delta\Theta_{Faraday}$ of $\sim 74^o$.
But it also has to be noted that any general Faraday rotation occurring between 
the shell and the observer would not affect the general distribution observed,
as it would rotate the overall polarisation vector angles by the same amount.

\item Guilain \& Mauron (1996) showed that for an Oxygen-rich AGB star a 
typical fractional electron abundance $x_{e} \sim 2 \times 10^{-5}$. Such a 
value would produce a maximum differential rotation of the polarisation 
vectors of $\sim 10^{\circ}$. This would not change our fundamental results.

\item  Internal Faraday rotation from the denser central region itself is 
expected to be negligible since the size of the thin ionized hydrogen layer 
surrounding the central star has a typical thickness of less than 10$^{12}$cm. 
The overall region size is typically less than 2R$_{star}$ (i.e., $<$3~AU) to 
be compared with the typical maser spot size of at least 10-20~AU. 
This means that in the most pessimistic scenario it would affect at the very 
most 30\% of the emission of those red-shifted maser spots on or very near the 
line of sight. 

\end{itemize}

Consequently, we exclude Faraday rotation as a possible cause for the dichotomy 
observed in the polarisation angle distribution.

\subsection{Shell structure and extent}

At 1667~MHz, the emission in the south and the north-east of the shell 
extends beyond that at 1612~MHz by nearly 1~arcsec. This is an interesting 
result since mainline emission (i.e., 1665 and 1667~MHz) is expected to be 
internal to that of the 1612~MHz satellite line due to the difference in 
pumping schemes. The 1612~MHz transition is largely pumped by absorption at 
35 and 53~$\mu$m radiation, whilst the mainlines are pumped by a radiative 
absorption from the ground state to  $^2 \Pi_{3/2} \, J=5/2$, followed by a 
collisional de-excitation, which requires a high density than the 1612~MHz 
mechanism (Gray 2007). Competitive gain (Field 1985) can affect the balance of 
the intensities in the mainlines. Observing 1667~MHz emission beyond
that of the 1612~MHz requires deviation from the standard model 
(Collison \& Nedoluha 1995). \\

Fong et al. (2002) imaged the circumstellar envelope of OH~26.5+0.6 in 
the $^{12}$CO~$J=1-0$ line. Their observations show a 
deconvolved source size of 8.8 $\times$ 5.5 arcsec$^2$. In order to account 
for the observed flux, they needed to include a second, more tenuous, AGB wind 
and conclude that up to 80\% of the CO flux comes from the unresolved 
superwind. \\

Chesneau et al. (2005) observed OH~26.5+0.6 at 8.7~$\mu$m with 
the VLTI. Their deconvolved image exhibits asymmetry resulting in a elliptical
shape with an axis ratio of 0.75 and a mean position angle (PA) of
95$^\circ \pm$6$^\circ$. 
The authors suggest that the flattened distribution
observed in the mid-infrared could be explained by either an equatorial
overdensity or a disk close to an edge-on configuration. 

Both the ellipticity and the mean position angle inferred by the latter 
authors relating to material close to the star, are in agreement with our 
new findings but at the OH maser location. \\

All this observational evidence shows that divergence from spherical symmetry
is already present and observable at different resolutions in the whole 
gaseous and dusty envelope of OH~26.5+0.6.
This provides us with strong evidence that, in this case, the onset of
asymmetry does indeed start as early as the late-AGB phase. The
two-step mechanism proposed by Sahai (2002) in which a high-speed
collimated outflow (in other words, an anisotropic superwind) would
carve an imprint within an intrinsically spherical AGB mass-loss
envelope could be in action.

\subsection{Role of the magnetic field in the shaping process ?}

From the magnetic field strength of 3.7~mG measured from Zeeman splitting 
(cf Section~\ref{Pol section}), 
we can infer the corresponding magnetic energy density 
$\epsilon_{\small B}$ at the location of the OH maser emission: \\

\begin{equation}
 \epsilon_{\small B}= B^{2}/(2\mu_{0}) = \frac{10^{-1}}{8 \pi} B_{\rm Gauss}^2 
   = 5.5 \times 10^{-8}{\rm J \, m}^{-3}
\label{Equation: Mag Energy Density}
\end{equation}
 And, we can compare it with the thermal and kinetic energy 
densities $\epsilon_{\small Thermal}$ and $\epsilon_{\small Kinetic}$ 
respectively.
According to the model of Goldreich \& Scoville (1976), at a distance 
r$\sim10^{16}$cm, the number density of hydrogen in the wind of an OH/IR star 
is typically $n_{H} = 10^{5}$~cm$^{-3}$ which is generally taken to be the 
typical distance for (mainline) OH maser emission. Nevertheless, we found 
that the maser extent at 1612~MHz attests to a radius of about 3500~AU, that is 
r$\sim 5.5 \simeq 10^{16}$~cm=875$\times$R$_{\rm star}$ in the model of 
Goldreich \& Scoville, for which $n_{H}$ drops to $10^{4}$~cm$^{-3}$. 
Adopting nonetheless a conservative value of $n_{H} = 10^{5}$~cm$^{-3}$, and 
T=100~K, leads to the following for the thermal and kinetic densities:\\

\begin{equation}
\epsilon_{\small Thermal}= \frac{3}{2} n_{H}kT \sim 2 \times 10^{-10}{\rm J \,m}^{-3}
\label{Equation: Thermal Energy Density}
\end{equation}

\noindent
and  \\

\begin{equation}
\epsilon_{\small Kinetic}=\frac{1}{2} \rho V_{exp}^2 
  \sim  1.8 \times 10^{-8} {\rm J \,m}^{-3}
\label{Equation: Kinetic Energy Density}
\end{equation}

\noindent
with  $V_{exp}$=15~km~s$^{-1}$. \\
 
This indicates that the magnetic energy density dominates over the thermal 
energy density and is at least 3 times greater than the kinetic energy 
density. \\

\begin{figure}
\noindent
{\bf a)} \\
\hspace*{-0.55cm} \epsfig{file=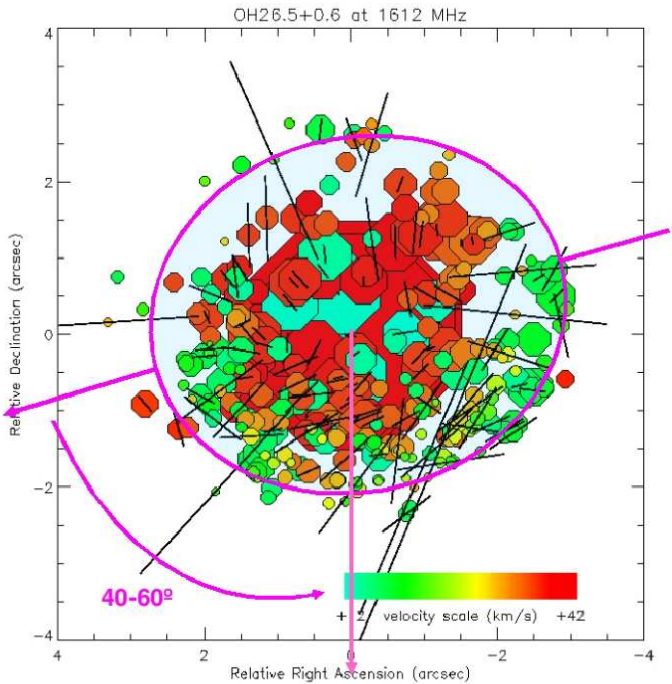,width=9.75cm,angle=0}
\noindent
{\bf b)} \\
\hspace*{-2.75cm} \epsfig{file=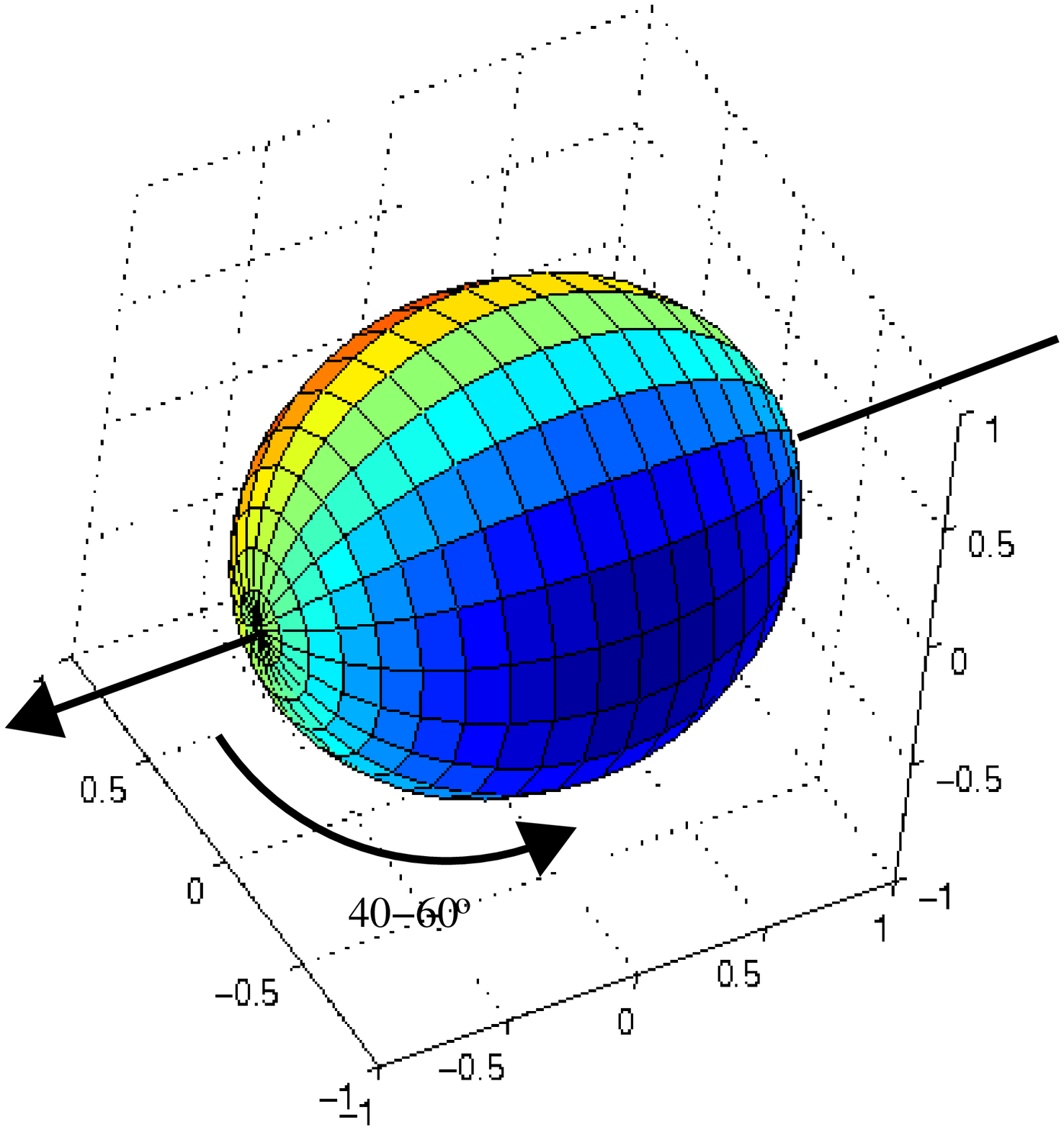,width=11.00cm,angle=-90}
   \caption{{\bf a)} Distribution of the maser spots at 1612~MHz as shown in 
       Fig.~\ref{fig: 1612 maser spot}, on which is overlaid a model explaining 
       both the geometric and polarimetric structures observed. {\bf b)} A 3D
       representation of the model in which the line of sight is perpendicular
       to the plane of the page.}
     \label{fig: maser spot distrib and model}
   \end{figure}

Figure~\ref{fig: maser spot distrib and model} presents, superimposed on top of 
the maser spot and polarisation vector distributions: 1) the ellipse that 
describes best the maser spot distribution observed; 2) the axis 
separating the radial and tangential vectors of polarisation, and 3) the
direction of the magnetic field which would produce such a distribution of the
vectors of polarisation.

As previously mentioned, the position angle determined by 
Chesneau et al. (2005) for the mid-infrared emission is in agreement with the 
semi-major axis orientation on the plane of the sky of the 1612~MHz ellipsoid 
distribution as observed here.

The axis separating the tangential and radial distribution of the polarisation 
vectors accounting for a magnetic field direction of 40-60$^{\circ}$ is 
aligned with the major axis of the geometrically ellipsoidal maser emission 
(cf. Fig.~\ref{fig: maser spot distrib and model}).
This latter projected ellipse, and the velocity distribution of the maser spots
as observed in Fig.~\ref{fig: vel dist and models} are indeed expected if the 
actual shell geometry is a prolate spheroid tilted about 45-65$^\circ$ to the
line of sight (Bowers 1991).
These combined results reveal that there is a definite correlation between the
magnetic field orientation and the geometrical structure of the circumstellar
envelope.

\section{Conclusion}

The infrared and 18~cm OH maser properties of OH~26.5+0.6 attest to a thick 
circumstellar envelope, characteristic of a rather evolved star most
probably at the tip of the AGB.
The 1612~MHz emission reveals an ellipsoidal geometry, while the presumably 
more central 1665~MHz emission traces a filamentary structure. Both 1612 and 
1667~MHz high resolution maps show a lack of maser emission from some parts of
the shell, and in the southern and north-eastern part of the shell, 1667~MHz
emission extends beyond that at 1612~MHz. 
All these deviations from the standard spherical model
show that the OH/IR stage (i.e., late-AGB phase) is clearly at the stage where 
asymmetry starts to develop.
The presence of acceleration in the shell at the OH maser location 
may be a secondary factor enhancing the asymmetry observed so far away from 
the star. The root of this asymmetry is likely to be close to the stellar 
surface itself, as near infrared results indicate. This latter hypothesis is 
reinforced by the agreement in orientation of the major axis of the elliptic 
distribution observed in infrared and at 1612~MHz. Finally, we found 
that the magnetic field strength, inferred from OH Zeeman splitting, is such 
that the magnetic field energy density dominates over the thermal and kinetic 
pressures and that there is a definite correlation between the magnetic field 
orientation and the main axis of geometrically ellipsoidal maser emission. 
This suggests that the magnetic field plays a role in the shaping 
process observed. 
\section*{Acknowledgements}
The authors would like to thank L.E. Davis who contributed in the primary data 
reduction during her stay as a visiting student at Jodrell Bank 
Observatory. 
We thank the referee for useful comments, some of which prompted us changes 
improving the clarity of the paper. 
We would also like to thank Malcolm D. Gray for his careful reading of the 
manuscript and for constructive discussions. 
Finally, we are grateful to Aris Noustos for the calculation of the RM 
applicable to OH~26.5+0.6 \\ 
The work presented here is based on observations obtained with MERLIN, a 
National Facility operated by the University of Manchester at Jodrell Bank 
Observatory, on behalf of STFC.  


\newpage
\appendix
\begin{table*}
\caption{Stokes parameter flux densities and polarisation properties 
         of the 1612~MHz maser spots.}
\label{Table: 1612 maser spot}
{\small
\begin{tabular}{rrrrrrrrrrrrr}
\hline
No            & Vel           & I             &  Q            & U             & V               & P             &
$\Delta \alpha$ & $\Delta \delta$ & $m_{\rm c}$   & $m_{\rm l}$       & $m_{\rm t}$       & $\chi$        \\
              & (km~s$^{-1}$) & (Jy~b$^{-1}$) & (Jy~b$^{-1}$) & (Jy~b$^{-1}$) & (Jy~b$^{-1}$)   & (Jy~b$^{-1}$) &
(milli arcsec)& (milli arcsec)& (\%)      & (\%)          & (\%)          & ({\degr})         \\
\hline   
    1 & 41.921 &  4.829 &  0.088 & -0.147 &  0.121 &  0.171 &   501.38 &  -236.10 &   2.5 &  3.5 &  4.3 & -29.5\\
    2 & 41.139 & 63.344 & -0.236 &  0.551 & -0.169 &  0.600 &    58.00 &   -16.10 &  -0.3 &  0.9 &  0.9 &  56.6\\
    3 & 40.873 & 50.169 & -0.394 &  0.416 & -0.520 &  0.573 &   -19.88 &   -16.16 &  -1.0 &  1.1 &  1.5 &  66.7\\
    4 & 40.676 &  5.481 &  0.138 & -0.013 & -0.340 &  0.139 &   869.41 &  -107.05 &  -6.2 &  2.5 &  6.7 &  -2.7\\
    5 & 40.344 &  1.750 & -0.004 & -0.020 & $<$-0.012 & $<$0.020 &  1399.67 &  -510.60 & $<$-0.7 & $<$1.2 & $<$1.4 & / \\
    6 & 40.123 &  2.518 &  0.030 &  0.003 &  0.023 &  0.030 &  -174.57 &   694.78 &   0.9 &  1.2 &  1.5 &   2.9\\
\end{tabular}
}
\end{table*}
        \addtocounter{table}{-1}%
\begin{table*}
\caption{ \it continued}
{\small
\begin{tabular}{rrrrrrrrrrrrr}
\hline
No            & Vel           & I             &  Q            & U             & V               & P             &
$\Delta \alpha$ & $\Delta \delta$ & $m_{\rm c}$   & $m_{\rm l}$       & $m_{\rm t}$       & $\chi$        \\
              & (km~s$^{-1}$) & (Jy~b$^{-1}$) & (Jy~b$^{-1}$) & (Jy~b$^{-1}$) & (Jy~b$^{-1}$)   & (Jy~b$^{-1}$) &
(milli arcsec)& (milli arcsec)& (\%)      & (\%)          & (\%)          & ({\degr})         \\
\hline   
   65 & 37.119 &  0.723 &  0.009 &  0.008 & $<$-0.004 &  $<$0.012 &  331.27 &  1447.99 & $<$-0.6 & $<$1.7 & $<$1.8 & / \\
   66 & 37.012 &  0.552 & -0.002 & -0.016 & -0.031 &  $<$0.016 &   170.30 &  -600.27 &  -5.6 &  $<$2.9 &  $<$6.3 & / \\
   67 & 37.002 &  0.446 & -0.001 &  0.000 & -0.040 &  $<$0.001 &  1919.21 &   -28.14 &  -9.0 &  $<$0.2 &  $<$9.0 & / \\
   68 & 36.936 &  0.411 &  0.005 &  0.004 & $<$-0.006 & $<$ 0.006 &   62.48 &  2260.37 & $<$-1.5 & $<$1.6 & $<$2.2 & / \\
   69 & 36.920 &  0.580 & -0.008 & -0.050 & -0.038 &  0.051 &    88.38 & -1646.85 &  -6.6 &  8.7 & 10.9 & -49.5\\
   70 & 36.880 &  0.657 & -0.006 & -0.009 & -0.040 &  $<$0.011 &  1233.32 &  -790.36 &  -6.1 &  $<$1.6 &  $<$6.3 & / \\
\end{tabular}
}
\end{table*}
        \addtocounter{table}{-1}%
\begin{table*}
\caption{ \it continued}
{\small
\begin{tabular}{rrrrrrrrrrrrr}
\hline
No            & Vel           & I             &  Q            & U             & V               & P             &
$\Delta \alpha$ & $\Delta \delta$ & $m_{\rm c}$   & $m_{\rm l}$       & $m_{\rm t}$       & $\chi$        \\
              & (km~s$^{-1}$) & (Jy~b$^{-1}$) & (Jy~b$^{-1}$) & (Jy~b$^{-1}$) & (Jy~b$^{-1}$)   & (Jy~b$^{-1}$) &
(milli arcsec)& (milli arcsec)& (\%)      & (\%)          & (\%)          & ({\degr})         \\
\hline   
  129 & 32.526 &  0.074 &  0.018 & -0.020 &  $<$0.006 &  0.027 & -862.02 & -2005.87 & $<$8.1 & 36.4 & $<$37.3 & -24.0\\
  130 & 32.519 &  0.093 & -0.023 & -0.003 &  $<$0.007 &  0.023 & 3315.37 &   159.03 & $<$7.5 & 24.9 & $<$26.0 & -86.3\\
  131 & 32.507 &  0.241 & -0.013 & -0.005 & $<$-0.016 &  $<$0.014 &-1346.97 & -1375.27 & $<$-6.6 & $<$5.8 & $<$8.8 & / \\
  132 & 32.474 &  0.101 & -0.025 &  0.005 &  $<$0.008 &  0.025 &-2241.52 &   258.98 & $<$7.9 & 25.2 & $<$26.4 &  84.3\\
  133 & 32.447 &  0.271 & -0.011 & -0.026 & $<$-0.011 & 0.028 &-1636.80 &  -970.94 & $<$-4.1 & 10.4 & $<$11.2 & -56.5\\
  134 & 32.276 &  0.085 &  0.010 & -0.006 & $<$-0.010 & $<$0.012 &2165.88 & -849.78 & $<$-11.8 & $<$13.7 & $<$18.1 & / \\
\end{tabular}
}
\end{table*}
        \addtocounter{table}{-1}%
\begin{table*}
\caption{ \it continued}
{\small
\begin{tabular}{rrrrrrrrrrrrr}
\hline
No            & Vel           & I             &  Q            & U             & V               & P             &
$\Delta \alpha$ & $\Delta \delta$ & $m_{\rm c}$   & $m_{\rm l}$       & $m_{\rm t}$       & $\chi$        \\
              & (km~s$^{-1}$) & (Jy~b$^{-1}$) & (Jy~b$^{-1}$) & (Jy~b$^{-1}$) & (Jy~b$^{-1}$)   & (Jy~b$^{-1}$) &
(milli arcsec)& (milli arcsec)& (\%)      & (\%)          & (\%)          & ({\degr})         \\
\hline   
  193 & 20.595 &  0.208 & -0.009 &  0.005 & $<$-0.003 &  $<$0.010 & 1647.09 &  -444.01 & $<$-1.4 & $<$4.9 & $<$5.1 & / \\
  194 & 20.470 &  0.670 &  0.005 &  0.004 & $<$-0.004 &  $<$0.006 & 2183.36 &  -398.83 & $<$-0.6 & $<$1.0 & $<$1.2 & / \\
  195 & 20.458 &  0.203 &  0.008 & -0.032 &  0.025 &  0.033 &   -61.33 & -1966.64 &  12.3 & 16.2 & 20.3 & -38.0\\
  196 & 20.456 &  0.222 &  0.001 & -0.004 &  $<$0.004 &  $<$0.004 & -321.62 & -1416.98 &  $<$1.8 & $<$1.9 & $<$2.6 & / \\
  197 & 20.440 &  0.138 &  0.000 & -0.004 & $<$-0.018 &  $<$0.004 &-2462.98 & 961.17 & $<$-13.0 & $<$2.9 & $<$13.3 & / \\
  198 & 20.424 &  0.575 &  0.006 & -0.005 & $<$ 0.022 &  $<$0.008 & 1933.90 &  -927.86 &  $<$3.8 & $<$1.4 & $<$4.0 & / \\
\end{tabular}
}
\end{table*}
        \addtocounter{table}{-1}%
\begin{table*}
\caption{ \it continued}
{\small
\begin{tabular}{rrrrrrrrrrrrr}
\hline
No            & Vel           & I             &  Q            & U             & V               & P             &
$\Delta \alpha$ & $\Delta \delta$ & $m_{\rm c}$   & $m_{\rm l}$       & $m_{\rm t}$       & $\chi$        \\
              & (km~s$^{-1}$) & (Jy~b$^{-1}$) & (Jy~b$^{-1}$) & (Jy~b$^{-1}$) & (Jy~b$^{-1}$)   & (Jy~b$^{-1}$) &
(milli arcsec)& (milli arcsec)& (\%)      & (\%)          & (\%)          & ({\degr})         \\
\hline   
  257 & 16.029 &  0.466 & -0.018 & -0.018 &  $<$0.006 &  0.025 & -1919.01 & -1176.00 &  $<$1.3 & 5.5 & $<$5.7 & -67.5\\
  258 & 15.968 &  0.799 & -0.009 &  0.010 & -0.039 &  $<$0.013 &  1652.97 &  -350.53 &  -4.9 &  $<$1.7 &  $<$5.2 & / \\
  259 & 15.948 &  0.450 & -0.019 & -0.004 & -0.041 &  $<$0.019 &  1082.06 & -2015.52 &  -9.1 &  $<$4.3 & $<$10.1 & / \\
  260 & 15.805 &  1.069 & -0.060 & -0.011 &  0.027 &  0.061 &   516.80 &  -790.25 &   2.5 &  5.7 &  6.2 & -84.8\\
  261 & 15.751 &  0.514 &  0.015 &  0.006 &  0.028 &  $<$0.016 &   -46.10 &  2614.48 &   5.4 &  $<$3.1 &  $<$6.2 & / \\
  262 & 15.685 &  0.770 & -0.009 & -0.006 & $<$-0.019 &  $<$0.011 &-2255.16 &  -584.81 & $<$-2.5 & $<$1.4 & $<$2.9 & / \\
\hline
\end{tabular}
}
\end{table*}


\begin{table*}
\caption{Stokes parameter flux densities and polarisation properties 
         of the 1667~MHz maser spots.}
\label{Table: 1667 maser spot}
{\small
\begin{tabular}{rrrrrrrrrrrrr}
\hline
No            & Vel           & I             &  Q            & U             & V               & P             &
$\Delta \alpha$ & $\Delta \delta$ & $m_{\rm c}$   & $m_{\rm l}$       & $m_{\rm t}$       & $\chi$        \\
              & (km~s$^{-1}$) & (Jy~b$^{-1}$) & (Jy~b$^{-1}$) & (Jy~b$^{-1}$) & (Jy~b$^{-1}$)   & (Jy~b$^{-1}$) &
(milli arcsec)& (milli arcsec)& (\%)      & (\%)          & (\%)          & (\degr)         \\
\hline     
    1 & 41.738 &  0.128 &  0.003 &  0.001 &  $<$0.001 &  $<$0.003 &  -408.58 &  1145.12 &   $<$0.8 &   $<$2.5 &   $<$2.6 & / \\
    2 & 41.578 &  0.058 &  0.001 &  0.002 &  $<$0.001 &  $<$0.002 &  -263.12 &  -292.51 &   $<$1.7 &   $<$3.9 &   $<$4.3 & / \\
    3 & 41.375 &  0.049 &  0.000 &  0.001 &  $<$0.002 &  $<$0.001 &  -445.48 &   -49.43 &   $<$4.1 &   $<$2.0 &   $<$4.6 & / \\
    4 & 41.289 &  0.066 &  0.000 & -0.001 & $<$-0.002 &  $<$0.001 &  -679.07 &   493.79 &  $<$-3.0 &   $<$1.5 &   $<$3.4 & / \\
    5 & 40.476 &  0.089 & -0.001 &  0.004 & $<$-0.004 &  $<$0.004 & -1112.70 &  -114.53 &  $<$-4.5 &   $<$4.6 &   $<$6.4 & / \\
    6 & 40.257 &  0.105 & -0.002 &  0.003 &  $<$0.003 &  $<$0.004 &  -798.36 &  1266.66 &   $<$2.9 &   $<$3.4 &   $<$4.5 & / \\
\end{tabular}
}
\end{table*}
        \addtocounter{table}{-1}%
\begin{table*}
\caption{ \it continued}
{\small
\begin{tabular}{rrrrrrrrrrrrr}
\hline
No            & Vel           & I             &  Q            & U             & V               & P             &
$\Delta \alpha$ & $\Delta \delta$ & $m_{\rm c}$   & $m_{\rm l}$       & $m_{\rm t}$       & $\chi$        \\
              & (km~s$^{-1}$) & (Jy~b$^{-1}$) & (Jy~b$^{-1}$) & (Jy~b$^{-1}$) & (Jy~b$^{-1}$)   & (Jy~b$^{-1}$) &
(milli arcsec)& (milli arcsec)& (\%)      & (\%)          & (\%)          & ({\degr})         \\
\hline   
   65 & 17.760 &  0.083 &  0.004 &  0.003 &  $<$0.006 &  $<$0.005 &  -831.41 & -1816.88 &   $<$7.2 &   $<$6.0 &   $<$9.4 & / \\
   66 & 17.550 &  0.044 &  0.006 & -0.007 &  $<$0.005 &  $<$0.009 &  -521.83 & -1955.23 &  $<$11.4 &  $<$21.0 &  $<$23.9 & / \\
   67 & 16.585 &  0.068 &  0.000 &  0.001 &  $<$0.010 &  $<$0.001 &  -932.56 &  1659.17 &  $<$14.7 &   $<$1.5 &  $<$14.8 & / \\
   68 & 16.438 &  0.098 &  0.000 & -0.001 & $<$-0.005 &  $<$0.001 &  -175.36 &  -746.33 &  $<$-5.1 &   $<$1.0 &   $<$5.2 & / \\
   69 & 16.331 &  0.056 &  0.002 & -0.002 & $<$-0.018 &  $<$0.003 & -1823.00 &  1254.70 & $<$-32.1 &   $<$5.1 &  $<$32.5 & / \\
   70 & 16.051 &  0.113 & -0.002 & -0.010 &  $<$0.002 &  $<$0.010 &  -806.40 &  -753.89 &   $<$1.8 &   $<$9.0 &   $<$9.2 & / \\
\hline
\end{tabular}
}
\end{table*}


\begin{table*}
\caption{Stokes parameter flux densities and polarisation properties 
         of the 1665~MHz maser spots.}
\label{Table: 1665 maser spot}
{\small
\begin{tabular}{rrrrrrrrrrrrr}
\hline
No            & Vel           & I             &  Q            & U             & V               & P             &
$\Delta \alpha$ & $\Delta \delta$ & $m_{\rm c}$   & $m_{\rm l}$       & $m_{\rm t}$       & $\chi$        \\
              & (km~s$^{-1}$) & (Jy~b$^{-1}$) & (Jy~b$^{-1}$) & (Jy~b$^{-1}$) & (Jy~b$^{-1}$)   & (Jy~b$^{-1}$) &
(milli arcsec)& (milli arcsec)& (\%)      & (\%)          & (\%)          & (\degr)         \\
\hline   
  1 & 41.31 & 0.088 &  0.007 &  0.005 &  0.023    & $<$0.009 &  1215.12 &  280.05 &  26.1    &  $<$9.8 & $<$27.9 &   /   \\
  2 & 41.22 & 0.068 &  0.006 & -0.005 &  $<$0.008 & $<$0.008 &   865.90 & 1498.96 &  $<$11.8 & $<$11.5 & $<$16.5 &   /   \\
  3 & 40.99 & 0.107 & -0.002 &  0.006 &  $<$0.008 & $<$0.006 &  -708.11 &  266.12 &   $<$7.5 &  $<$5.9 &  $<$9.5 &   /   \\
  4 & 40.85 & 0.055 &  0.003 &  0.007 &  $<$0.001 & $<$0.008 &  -411.86 & -239.71 &   $<$1.8 & $<$13.8 & $<$13.9 &   /   \\
  5 & 40.78 & 0.093 & -0.008 & -0.007 &  $<$0.013 & $<$0.011 &  -218.68 & -403.23 &  $<$14.0 & $<$11.4 & $<$18.1 &   /   \\
  6 & 40.66 & 0.109 & -0.010 &  0.008 & $<$-0.006 & $<$0.013 & -1240.89 &  743.75 &  $<$-5.5 & $<$11.7 & $<$12.9 &   /   \\
  7 & 40.35 & 0.091 & -0.008 & -0.015 & $<$-0.018 & $<$0.017 & -1406.84 & 1042.45 & $<$-19.8 & $<$18.7 & $<$27.2 &   /   \\
  8 & 40.24 & 0.085 &  0.010 & -0.012 & $<$-0.003 & $<$0.016 & -1926.92 & 1717.89 &  $<$-3.5 & $<$18.4 & $<$18.7 &   /   \\
  9 & 12.01 & 0.583 &  0.040 & -0.034 &  $<$0.007 & 0.052    &     4.38 &  -17.37 &   $<$1.2 &  9.0 &  9.1 & -20.2 \\
 10 & 11.87 & 0.420 &  0.030 & -0.007 &  0.016    & 0.031    &   -22.98 &  114.11 &   3.8    &  7.3 &  8.2 &  -6.6 \\
\hline
\end{tabular}
}
\end{table*}



\begin{thebibliography}{}
\bibitem[\protect\citename{Andersson etal}%
 1974]{Andersson1974} Andersson C., Johansson L.E.B., Goss W.M., 
 Winnberg A., Nguyen-Quang-Rieu, 1974, A\&A, 30, 475
\bibitem[\protect\citename{Bains etal}%
 2003]{Bains03} Bains I., Gledhill T.M., Yates J.A., Richards, A.M.S.,
  2003, MNRAS, 338, 287
\bibitem[\protect\citename{Baud}%
 1981]{Baud1981} Baud B., 1981, ApJ, 250L, 79
\bibitem[\protect\citename{Boboltz}%
 1997]{Boboltz1997} Boboltz D.A., 1997, PhD
\bibitem[\protect\citename{Bowers etal.} 
 1983]{Bowers83} Bowers P.F., Johnston K.J., Spencer J.H., 1983, ApJ, 274, 733
\bibitem[\protect\citename{Bowers and Johnston 1990}%
 1990]{Bowers90} Bowers P.F., Johnston K.J., 1990, ApJ, 354, 676
\bibitem[\protect\citename{Bowers 1991}%
 1990]{Bowers91} Bowers P.F., 1991, ApJS, 76, 1099
\bibitem[\protect\citename{Chesneau et al}%
 2005]{Chesneau05} Chesneau O., Verhoelst T., Lopez B., Waters L.B.F.M., 
 Leinert Ch., Jaffe W., K\"{o}hler R., de~Koter A., Dijkstra C., 2005, 
 A\&A, 435, 563
\bibitem[\protect\citename{Collison}%
 1995]{Collison95} Collison A.J., Nedoluha G.E., 1995, ApJ, 442, 311
\bibitem[\protect\citename{Cordes01}%
 2001]{Cordes01} Cordes J. M., Lazio T. J. W., 2002, astro.ph, 7156
\bibitem[\protect\citename{Corradi}%
 1995]{Corradi95} Corradi R.L.M., Schwarz H.E., 1995, A\&A, 293, 871
\bibitem[\protect\citename{Diamond etal}%
 1985]{Diamond1985} Diamond P.J., Norris R.P., Rowland P.R., Booth R.S., 
 Nyman L-A., 1985, MNRAS, 212, 1
\bibitem[\protect\citename{Elitzur96}%
 1992]{Elitzur96} Elitzur M., 1996, ApJ, 457, 415
\bibitem[\protect\citename{Etoka, LeSqueren 00}%
 2004]{Etoka2000} Etoka S., Le~Squeren A.M., 2000, A\&AS, 146, 179
\bibitem[\protect\citename{Etoka, Diamond}%
 2004]{Etoka2004a} Etoka S., Diamond P.J., 2004, MNRAS, 348, 34 (paper~I)
\bibitem[\protect\citename{Etoka, LeSqueren 04}%
 2004]{Etoka2004b} Etoka S., Le~Squeren A.M., 2004, A\&A, 420, 217
\bibitem[\protect\citename{Field}%
 1995]{Field85} Field D., 1985, MNRAS, 217, 1
\bibitem[\protect\citename{Fond02}%
 2002]{Fong02} Fong D., Justtanont K., Meixner M., Campbell M. T., 2002, 
 A\&A, 396, 581 
\bibitem[\protect\citename{Goldreich73}%
 1973]{Goldreich76} Goldreich P., Keeley D. A., Kwan J. Y., 1973, ApJ, 
 179, 111 
\bibitem[\protect\citename{Goldreich76}%
 1976]{Goldreich76} Goldreich P., Scoville N., 1976, ApJ, 205, 144
\bibitem[\protect\citename{Gray07}%
 2007]{Gray07} Gray M.D., 2007, MNRAS, 375, 477
\bibitem[\protect\citename{Guilain96}%
 1996]{Guilain96} Guilain C., Mauron N., 1996 A\&A, 314, 585
\bibitem[\protect\citename{Habing96}%
 1985]{Habing96} Habing H., 1996, A\&A Rev, 7, 97 
\bibitem[\protect\citename{Harvey etal}%
 1985]{Harvey74} Harvey P.M., Bechis K.P., Wilson W.J., Ball J.A., 
 1974, ApJS, 27, 331
\bibitem[\protect\citename{Herman etal}%
 1985]{Herman85}  Herman J., Baud B., Habing H.J., Winnberg A., 1985, 
 A\&A, 143, 122
\bibitem[\protect\citename{Justtanont etal.}%
 1996]{Justtanont96} Justtanont K., Skinner C.J., Tielens A.G.G.M., 
 Meixner M., Baas F., 1996, ApJ, 456, 337  
\bibitem[\protect\citename{Kemball 97}%
 1996]{Kemball97} Kemball A.J., Diamond P.J., 1997, ApJ, 481, L111  
\bibitem[\protect\citename{Kemball 09}%
 2009]{Kemball09} Kemball A.J., Diamond P.J., Gonidakis I., Mitra M., 
 Yim K., Pan K.-C., Chiang H.-F., 2009, ApJ, 698, 1721  
\bibitem[\protect\citename{Le Bertre}%
 1993]{LeBertre93} Le~Bertre T., 1993, A\&AS, 97, 729
\bibitem[\protect\citename{Nordhaus etal}%
 2007]{Nordhaus07} Nordhaus J., Blackman E.G., Frank A., 2007, MNRAS, 376, 599
\bibitem[\protect\citename{Norris etal}%
 1984]{Norris84} Norris R.P., Booth, R.S., Diamond P.J., Nyman L.-A., Graham
 D. A., Matveenko L. I., 1984, MNRAS, 208, 435
\bibitem[\protect\citename{Noutsos}%
 2008]{Noutsos08} Noutsos A., Johnston S., Kramer M., 
 Karastergiou A., 2008, MNRAS, 386, 1881	
\bibitem[\protect\citename{Ortiz et al}%
 2006]{Ortiz06} Ortiz R., Lorenz-Martins S., Maciel W.J., Rangel E.M., 2005,
 A\&A, 431, 565 
\bibitem[\protect\citename{Reid etal}%
 1977]{Reid77} Reid M.J., Muhleman D.O., Moran J.M., Johnston K.J., 
 Schwartz P.R., 1977, ApJ, 214, 60
\bibitem[\protect\citename{Sahai etal}%
 2002]{Sahai02} Sahai R., 2002, \emph{Rev. Mex. Astron. Astrofis. Ser. Conf.}
  13, 133
\bibitem[\protect\citename{Savenster}%
 2002]{Sevenster02} Sevenster M., 2002, AJ, 123, 2772
\bibitem[\protect\citename{Sivagnanam etal}%
1990] {Sivagnanam90} Sivagnanam P., Diamond P.J., Le Squeren A.M., 
 Biraud F., 1990, A\&A, 229, 171 
\bibitem[\protect\citename{vanLangevelde etal 1990}%
 1990]{vanLangevelde90} van~Langevelde H.J., van~der~Heiden R., 
 van~Schooneveld C., 1990, A\&A, 239, 193
\bibitem[\protect\citename{vanLangevelde etal 2000}%
 2001]{vanLangevelde00} van Langevelde H.J., Vlemmings W., Diamond P. J.,
 Baudry A., Beasley A.J., 2000, A\&A, 357, 945
\bibitem[\protect\citename{Vlemming etal 2005}%
 2005]{Vlemmings05} Vlemmings V.H.T., van~Langevelde H.J., Diamond P.J., 2005, 
 A\&A, 434, 1029
\bibitem[\protect\citename{Vlemming et Diamond 2006}%
 2006]{Vlemmings06} Vlemmings V.H.T., Diamond P.J., 2006, ApJ 648L, 59
\end{thebibliography}
\end{document}